\documentclass[10pt,journal,compsoc]{IEEEtran}
\ifCLASSOPTIONcompsoc
  \usepackage[nocompress]{cite}
\else
  \usepackage{cite}
\fi
\usepackage{url}
\usepackage{amsmath,amssymb,amsfonts}
\usepackage{algorithmic}
\usepackage{graphicx}
\usepackage{textcomp}
\usepackage{xcolor}
\usepackage{tikz}
\usepackage{hyperref}
\usepackage[noend]{algorithm2e}
\usepackage{tabularx,booktabs}
\def\BibTeX{{\rm B\kern-.05em{\sc i\kern-.025em b}\kern-.08em
    T\kern-.1667em\lower.7ex\hbox{E}\kern-.125emX}}

\begin{document}

\title{Untargeted Poisoning Attack Detection in Federated Learning via Behavior Attestation}

\author{Ranwa~Al~Mallah,
        David~López, Godwin~Badu~Marfo, and~Bilal~Farooq
\IEEEcompsocitemizethanks{\IEEEcompsocthanksitem R. Al Mallah, G. Badu Marfo and B. Farooq are with the Laboratory of Innovations in Transportation, Ryerson University, Toronto, ON, Canada M5G1G3.\protect\\
E-mail: ranwa.almallah@ryerson.ca, gbmarfo@gmail.com, bilal.farooq@ryerson.ca
\IEEEcompsocthanksitem D. López is with the  Grupo de Investigación en Ingeniería de Transporte y Logística, Instituto de Ingeniería, Universidad Nacional Autónoma de México, Mexico City, Mexico, 04510.\protect\\
 E-mail: dlopezfl@iingen.unam.mx}
}
\markboth{}%
{Al Mallah \MakeLowercase{\textit{et al.}}: Untargeted Poisoning Attack Detection in Federated Learning via Behavior Attestation}

\IEEEtitleabstractindextext{
\begin{abstract}
Federated Learning (FL) is a paradigm in Machine Learning (ML) that addresses data privacy, security, access rights and access to heterogeneous information issues by training a global model using distributed nodes. Despite its advantages, there is an increased potential for cyberattacks on FL-based ML techniques that can undermine the benefits. Model-poisoning attacks on FL target the availability of the model. The adversarial objective is to disrupt the training. We propose attestedFL, a defense mechanism that monitors the training of individual nodes through state persistence in order to detect a malicious \textit{worker}. A fine-grained assessment of the history of the \textit{worker} permits the evaluation of its behavior in time and results in innovative detection strategies. We present three lines of defense that aim at assessing if the \textit{worker} is reliable by observing if the node is truly training, while advancing towards a goal. Our defense exposes an attacker's malicious behavior and removes unreliable nodes from the aggregation process so that the FL process converge faster. 
attestedFL increased the accuracy of the model in different FL settings, under different attacking patterns, and scenarios e.g., attacks performed at different stages of the convergence, colluding attackers, and continuous attacks.
\end{abstract}

}

\maketitle

%
\IEEEpeerreviewmaketitle

\IEEEraisesectionheading{
\section{Introduction}
\label{introduction}
}

\IEEEPARstart{F}{ederated} Learning (FL) has started to transform a number of industries. This new paradigm in Machine Learning (ML) addresses critical issues such as data privacy, access rights, security, and access to heterogeneous information by training a global model using distributed nodes \cite{konevcny2016federated}. The approach stands in contrast to the traditional centralized ML techniques, where all data samples are uploaded to a single server, as well as to more classical decentralized approaches, the local data samples are assumed to be identically distributed. Without the need for data sharing, FL trains an algorithm across multiple decentralized nodes that hold their local data samples.


Despite the advantages and successful application of FL in certain industry-based cases, there is an increased potential for cyberattacks on this ML technique, which can seriously undermine its benefits. In fact, by design, FL requires frequent communication between nodes during the learning process to be able to exchange the parameters of the ML model. This represents an attack surface from which an attacker can exploit system vulnerabilities to gain access to the learning process.


Regarding the cybersecurity of FL, the poisoning attack is an attack type that takes advantage of the ML model during training. Some poisoning attacks target the model integrity (backdoor attacks), while others target the ML model availability. In the first category, also known as targeted attacks, a backdoor is inserted so that the model's boundary shifts in some way as to include the malicious training data, thus compromising the integrity of the model. In fact, FL raises the challenge that hiding training data allows the attackers to inject backdoors into the global model \cite{gu2019badnets}. For instance, the attacks can modify an image classifier so that it assigns an attacker-chosen label to images with certain features, or force a word predictor to complete certain sentences with an attacker-chosen word \cite{bagdasaryan2018backdoor}. 

In the category of poisoning attacks that target the ML model availability also known as untargeted attacks, the aim is to generate poisoned local model updates to inject into the system so that the parameters that the global model learns become essentially useless. Model poisoning exploits the fact that FL gives malicious nodes direct and unrestricted influence over the global model. In fact, each round of the iterative process consists of training local models on local nodes to produce a set of potential model updates at each node. Thus, an attacker can impact the weights in the global model so that the new learned model never converges on the federated learning task. 


In the gamification process of attacks and defenses against untargeted attacks, many known defenses have shown that robust aggregators are provably effective under appropriate assumptions at mitigating untargeted attacks \cite{abadi2016tensorflow, blanchard2017machine, chen2017distributed, mhamdi2018hidden}. However, Fang et al. \cite{fang2020local} proposed a directed deviation attack that multiple aggregators did little to defend against. The existing defenses leave a margin for poisoning since the theoretical guarantees of these defenses may only hold under assumptions on the learning problem that are often not met \cite{kairouz2019advances}. Pan et al. \cite{pan2020justinian} proposed a defense where a reinforcement learning agent learns, from the updates provided by \textit{workers} and the relative decrease of loss as its reward, to propose the weight update of the global model at each iteration. In the approach, each coordinate of the learned policy is interpreted as the current credit on the corresponding node and attests to the reliability of the \textit{worker}. The authors proposed a detection technique, which consists in removing unreliable nodes at every iteration based only on the credits in the learned policy sequence.

Based on the open challenge to protect against untargeted attacks, in this work, we conduct an extensive detection and behavioral pattern analysis and propose \emph{attestedFL}, a defense mechanism that protects the system and can restore robustness in adversarial settings. While the untargeted attack tries to decrease the convergence of the model, attestedFL studies the behavior of the \textit{workers} over time through state persistence and removes unreliable nodes from the aggregation process so that the FL process converge faster. This results in innovative detection strategies. We present three lines of defense that aim at assessing if the \textit{worker} is reliable by observing if the node is truly training and advancing towards a goal. 


We conduct untargeted attacks on various federated learning settings and present different attacking patterns. We evaluate the impact of our defense and report the accuracy the model reaches epoch after epoch. The results of our extensive simulation analysis of different cyberattack scenarios show an increase in accuracy, demonstrating the efficiency and security of our proposed defense. Unlike the existing defenses, our proposed approach does not bound the expected number of malicious \textit{workers} and is shown to be robust in more challenging scenarios. 


\indent The contributions of this paper are summarized as follows:
\begin{itemize}
\item We propose the design of attestedFL, a fine-grained assessment of a \textit{worker's} behavior over time through state persistence for detection of untargeted attacks in federated learning, which does not require any upper bound on the malicious nodes ratio. Nodes that are not training represent unreliable \textit{workers} that must be eliminated. 
\item We present three lines of defense to protect against local model poisoning attacks. attestedFL-1 observes the convergence in time of the local model towards the global model. attestedFL-2 monitors the angular distance of successive local model updates throughout the training of a node and attestedFL-3 removes local model updates from \textit{workers} whose performance does not improve compared to their own previous performance on a \textit{quasi-validation dataset}.
\item We implement untargeted attacks and evaluate the impact of the defense in different FL settings and under different attacking patterns in order to validate the efficiency and security of our detection mechanism. We evaluate the scalability and computation overhead incurred by attestedFL.
  
\end{itemize}\vspace{1mm}

The rest of the paper is organised as follows. In Section~\ref{sec:realted_work} we present related work. In Section~\ref{sec:threat_model} we describe the threat model. We present attestedFL in Section~\ref{sec:attestedFL_design}. In Section~\ref{sec:evaluation}, we provide experimental results and analysis of the impact of the defense against poisoning attacks. We conclude the paper and provide future work in Section~\ref{sec:conclussions}.

\section{Related Work}

\label{sec:realted_work}
In this paper, we focus on the most common implementation of federated learning algorithms i.e. the data-parallel distributed learning system with one node acting as the \textit{chief} and a set of nodes acting as \textit{workers}. Although federated learning enables nodes to construct a machine learning model without sharing their private training data with each other, the fact that the \textit{chief} has no visibility into how the model updates are generated represents a major vulnerability in the process. Attacks exploit this vulnerability and when the adversarial goal is to ensure that the distributed implementation of the Stochastic Gradient Descent (SGD) algorithm converges to sub-optimal models, we consider that those model-poisoning attacks target the availability. The aim of the defense in this case is to ensure convergence to the true optima.


In the field of defending against Byzantine adversaries, approaches have been proposed to ensure the robust aggregation of distributed SGD against adversarial \textit{workers} sending poisoned gradients during the training phase \cite{bhagoji2018analyzing}. A very commonly used aggregation rule is averaging \cite{abadi2016tensorflow}. Byzantine \textit{workers} propose vectors farther away from the correct area of search. They can force the \textit{chief} to choose a vector that is too large in amplitude or too far in direction from the other vectors. Thus, the linear combinations give the adversary full control of the aggregated gradient \cite{blanchard2017machine}. Hence, Byzantine-resilient gradient aggregation rule techniques have been proposed to guarantee the production of gradients that will make the SGD process converge despite the presence of a minority of adversarial \textit{workers}. 

Among the Byzantine-resilient aggregation rule techniques, considering that the geometric median of means of independent and identically random gradients converges to the underlying gradient function, Chen et al. \cite{chen2017distributed} proposed an approach to select among the proposed vectors, the vector closest to a center. They take the vectors that minimize the sum of the squared distances to every other vector. This goes from linear to squared-distance-based aggregation rule. However, two Byzantine \textit{workers} can collude, one helping the other to be selected, by moving the center of all the vectors farther from the ''correct area''. The majority-based approach considers every subset of vectors, and identifies the subset with the smallest diameter. Combining the intuitions of the majority-based and the squared-distance-based methods, Blanchard et al. \cite{blanchard2017machine} proposed to choose the vector that is the closest to its neighbors. 

However, Mhamdi et al. \cite{mhamdi2018hidden} discovered a hidden vulnerability of distributed learning in Byzantium. They put forward that convergence is not enough and that in high number of dimensions, an adversary can build on the loss function's non-convexity to make SGD converge to ineffective models. Precisely, they prove that existing Byzantine-resilient schemes leave a margin of poisoning. This is due to the fact that each gradient aggregation rule technique performs a linear combination of the selected gradients. Thus, the final aggregated gradient might have one unexpectedly high coordinate. Depending on the learning rate, updating the model with such gradient may push and keep the parameter vector in a sub–space rarely reached with the usual, Byzantine–free distributed setup, thus offering sub–optimal models. 

Moreover, the assumptions made in the techniques for Byzantine-tolerant distributed learning are explicitly false for federated learning with adversarial nodes. Particularly, they assume that the nodes' training data are i.i.d. (independent and identically distributed), unmodified, and equally distributed \cite{bagdasaryan2018backdoor}. In reality, participants' local training datasets in federated learning are relatively small and drawn from different distributions. Since non-i.i.d. data are used in FL, each local model may be quite different from the global model. Thus, there are significant differences between the weights of individual models. 

Clustering techniques aim to detect model updates that are very different from what they should be. However, the attacker might generate poisoning models that are very similar to the true distribution (called ``inliers''), but that can still successfully mislead the model. Anomaly detectors consider the magnitudes of model weights (e.g., Euclidean distances between them) or measure cosine similarity between submitted models and the joint model. However, in federated learning, local training datasets on different devices may not be independently and identically distributed (i.e., non-IID) \cite{mcmahan2017communication}. Since FL take advantage of the diversity of participants with non-i.i.d. training data, including unusual or low-quality local data and by design, the \textit{chief} should accept even local models that have low accuracy and significantly diverge from the current global model. As the local training datasets on different \textit{worker} devices are more non-IID, the local models are more diverse, leaving more room for attacks. 

In terms of the attacks, back-gradient optimization based attack is the state-of-the-art untargeted data poisoning attack for multiclass classifiers. The Gaussian attack randomly crafts the local models on the compromised \textit{worker} devices. It was demonstrates that both attacks cannot effectively attack Byzantine-robust aggregation rules \cite{munoz2017towards}. However, Fang et al.~\cite{fang2020local} proposed a directed deviation attack and applied it to four Byzantine-robust federated learning methods. Their attack was successful and increased the error rates of the models although they were claimed to be robust against Byzantine failures. They craft local models for the compromised \textit{workers} to deviate a global model parameter the most towards the inverse of the direction along which the global model parameter would change without attacks. They proposed updates to make the \textit{j$^{th}$} global model parameter change direction (increases or decreases) upon the previous iteration. They also present two defenses to defend against their directed deviation attack. One generalized defense removes the local models that have large negative impact on the error rate of the global model (inspired by Reject on Negative Impact (RONI) that removes training examples that have large negative impact on the error rate of the model \cite{barreno2010security}), while the other defense removes the local models that result in large loss (inspired by TRIM that removes the training examples that have large negative impact on the loss \cite{jagielski2018manipulating}). However, their results show that the defenses are not effective enough, highlighting the need for new defenses against their directed deviation attack. Since their attack manipulates the local models in each iteration not for the purpose of training, this behavior can be detected by our approach because we aim at detecting unreliable nodes that are not training. 

Attackers craft local model updates by looking at the cross-sectional status of local model updates at each iteration. No detection strategy thus far forced the attackers to consider their behavior over time before crafting their attack. Our detection performs a longitudinal status monitoring of the actions of all nodes and thus forces attackers to craft models that have to be undetected from this perspective too. The attacker thus faces two challenges. First, the aggregation rules where the detection monitors each iteration at a time to exclude malicious updates by comparing \textit{workers} to each other, and second, our detection aims at further monitoring the history of a \textit{worker's} local model updates to reject the \textit{worker} if it is not training.

The challenge still remains to defend against untargeted model poisoning attacks and reduce the adversarial leeway that causes drift to sub–optimal models. Existing defense mechanisms omit inspecting the behavior in time of a particular node during its training process in order to detect an anomaly and reject poisonous model updates. Without temporal and dynamic monitoring methods, the \textit{chief} cannot detect and remove malicious or unreliable \textit{workers} from the system. 

\section{Threat Model}
\label{sec:threat_model}

We consider a federated learning process where the goal of the attacker is to conduct sabotage activities with the aim to disrupt the system.

\textbf{Attacker's Knowledge and Capability.} Among all nodes, we assume that the attacker controls the entire training process for one or a few compromised \textit{workers} and cannot observe the local data samples of other \textit{workers}. The attacker does not control the aggregation algorithm used to combine \textit{workers}' updates into the joint model, nor any aspects of the benign participants' training. We assume that benign nodes create their local models by correctly applying the training algorithm prescribed by federated learning to their local data. (1) The attacker controls the training dataset of any compromised \textit{worker}; (2) it controls the local training process and the hyperparameters; (3) it can modify the weights of the local model update and relay false information on behalf of the \textit{worker} when communicating with the \textit{chief} node; and, (4) it can change its local training from round to round.

Depending on the type of attack, the attacker may or may not need to access the local data samples depending on how the local models sent from the compromised \textit{worker} devices are crafted. We consider two cases depending on whether the attacker needs the local training datasets on the compromised \textit{workers} in order to fabricate the malicious local model update. If the attacker can craft the local models on the compromised \textit{worker} devices by directly modifying the weights in the local model updates sent back to the \textit{chief} such that the global model deviates the most towards the inverse of the direction along which the before-attack global model would change, then in this untargeted attack, the attacker didn't need to observe the local data samples of the compromised \textit{worker}. Untargeted attacks make the learnt model unusable and eventually lead to denial-of-service attacks \cite{fang2020local}. The attacker was able to target the availability and the aim of the defenses in this case is to ensure convergence.

However, if the attacker’s objective is to cause the jointly trained global model to misclassify a set of chosen inputs with high confidence, i.e., it seeks to poison the global model in a targeted manner as in \cite{bhagoji2019analyzing,bagdasaryan2020backdoor}, then for this attack type, the attacker would need to observe the local training dataset on the compromised \textit{worker} device in order to craft the malicious local model and target the integrity of the training.

\textbf{Attack.} We consider a poisoning attack where the adversary can leverage its knowledge of the submitted weights updates to send via adversarial \textit{workers}, malicious updates aiming at significantly decreasing convergence and preventing the distributed learning process from converging to a satisfying state. We do not make any assumptions on the true data distribution so the final model parameters are not known.

The attacker tries to deviate from the Global Model $GM^{t + 1}$ with an arbitrary Malicious Model $MM^t$:
\begin{equation}
\begin{split}
	GM^{t + 1} & = GM^t + \frac{r}{n}\sum\limits_{j=1}^{m} (LM_j^{t + 1} - GM^t) \\
	\Rightarrow MM^t & = GM^t + \frac{r}{n}\sum\limits_{j=1}^{m} (LM_j^{t + 1} - GM^t)
\end{split}
\end{equation}

where, $t$ is the current time slot, $r$ is the learning rate, $LM_j^{t+1}$ is a Local Model of \textit{worker} \textit{j} at time \textit{t}, $n$ is total number of nodes, and $m$ is the subset of \textit{workers} participating in the training. Since non-i.i.d. data is used in this case, each local model may be quite different from the global model. With the convergence of the global model, the deviations cancel out bit by bit, which can be denoted as:
\begin{equation} 
\label{eq:convergence}
\sum\limits_{j=1}^{m-1} (LM_j^{t + 1} - GM^t) \rightarrow 0 
\end{equation}

Based on this, an adversary may upload a model as:
\begin{equation}
\begin{split}
LM_m^{t + 1}  & = \frac{n}{r}MM^t - (\frac{n}{r}-1)GM^t -\\
              & \quad \sum\limits_{j=1}^{m-1} (LM_j^{t + 1} - GM^t) 
\end{split}
\end{equation}
by applying basic algebra and Equation~\ref{eq:convergence} then for a big enough $t$, $LM_m^{t + 1}$ can be defined as in Equation~\ref{eq:local_attack}: 
\begin{equation}
\label{eq:local_attack}
LM_m^{t + 1} \simeq \frac{n}{r}(MM^t - GM^t) + GM^t 
\end{equation}

\begin{figure}[b]
\begin{center}
  \includegraphics[width=1\columnwidth]{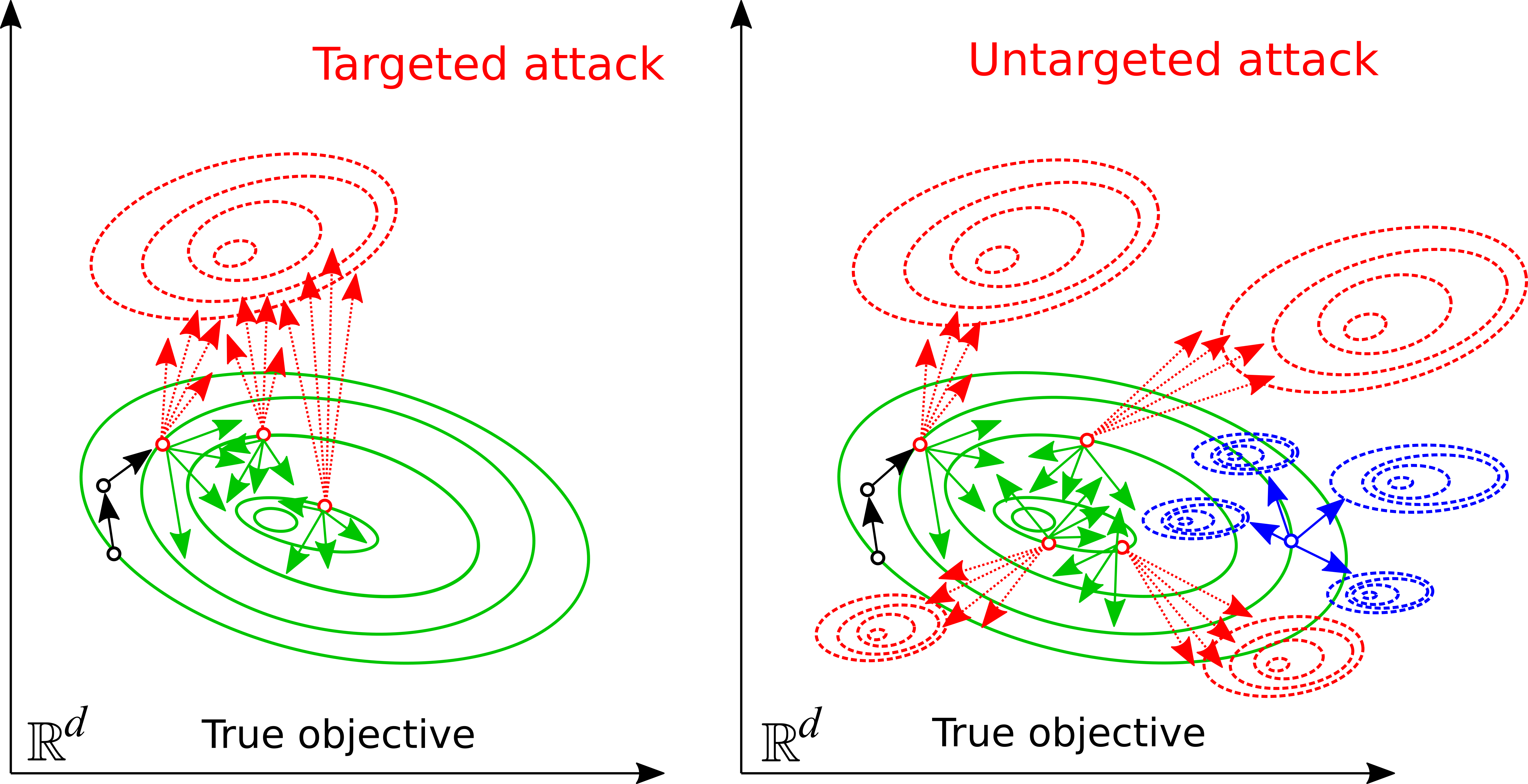}
  \caption{Model poisoning attacks - The dotted red arrows represent colluding attackers. The green circles represent the different stages of convergence of the global model. The green arrows represent benign \textit{worker} nodes. The blue arrows represents single attackers attacking individually towards different goals.}
  \label{fig:Attack_vis}
\end{center}
\end{figure}

We illustrate in Figure \ref{fig:Attack_vis} the targeted and untargeted attacks. If $MM^t$ remains the same at every time step, the attack tries to guarantee the replacement of the global model \textit{GM} by \textit{MM} and represents a targeted attack. In this case, attackers collude and together every epoch they go towards the same goal with the direction slightly changing as time advances. This is represented in the left diagram by the dotted red arrows pointing to the same red circle at every stage of convergence. In an untargeted attack, $MM^t$ is different from epoch to epoch, we show two scenarios in the figure. In the first scenario, one attacker may attack individually and on each epoch, goes in a different direction, towards different goals. This is represented in the right diagram by the blue arrows. In the second scenario, attackers may collude and together they go in a direction that is different from epoch to epoch i.e. each epoch different goals. This untargeted attack is represented in the right diagram by the dotted red arrows pointing to the different red circle at every stage of convergence. 

In targeted attacks, as time advances attackers are training, meaning that they are advancing towards the same goal. Whereas, in untargeted attacks, as time advances attackers are not training, they are not advancing towards a goal. Red arrows highlight this behavior. Targeted attacks are more costly for the attacker because the attacker will have to participate in the FL process at every iteration by locally training the model with its poison set. The steps of fabricating a poison set and training locally a model are not required in an untargeted attack. Although untargeted attacks are simpler to perform than targeted attacks, defense techniques against targeted attacks based on resilient gradient aggregation rules and clustering techniques do not remediate the untargeted scenario \cite{fang2020local}. 

\section{attestedFL Design}
\label{sec:attestedFL_design}

Our defense protects against untargeted attacks by reducing the drift towards sub-optimal models that they cause. Attackers in this context represent unreliable \textit{workers} that are not training and must be neutralized. During the training process of a \textit{worker}, there exists a complicated probabilistic dependency among the iterates. Such dependency cannot be specified from the distributed computing point of view. Since non-i.i.d. data is used in FL, each local model may be quite different from the global model. Thus, there are significant differences between the weights of individual models. However, a fine-grained assessment of the history of the \textit{worker} permits the evaluation of its behavior in time and results in innovative detection strategies. 

attestedFL uses three lines of defense to protect against untargeted local model poisoning attacks. The detection strategies aim at assessing if the \textit{worker} is reliable by observing if the node is really training iteration after iteration. Malicious nodes are not concerned in truly training their local model, but rather fabricating update values, thus they are considered unreliable. attestedFL removes the local models that are unreliable before computing the global model in each iteration of federated learning.

The first defense used by attestedFL is called attestedFL-1 and it monitors the history of the \textit{worker'} updated local models and observes the convergence of the local model towards the global model in order to remove the local models of \textit{workers} that appear not to be training. Many known defense mechanisms use the Euclidean distance function between two vectors $\parallel X - Y \parallel$ to provide byzantine resilient aggregation rules in order to transform the workers’ submitted gradients into a single, aggregated one \cite{mhamdi2018hidden,blanchard2017machine}. The idea of the mechanisms is to compare the Euclidean distance between model updates sent by different workers so that the chosen vector becomes the result of a distance minimization scheme. However, our reason for using Euclidean distance is to compare the Euclidean distance of consecutive model updates sent by the same worker. By doing so, we are able to examine the behaviour of the worker throughout the training in order to remove abnormal training behavior. By monitoring the history of each worker, we can remove the local models of workers that appear to be not training.

The second defense, attestedFL-2, monitors the angular distance of successive local model updates throughout the training of a node to remove abnormal training behavior. Some defense mechanisms in the literature also use the cosine similarity metric to detect malicious workers. For instance, the approach in \cite{fung2018mitigating} uses the cosine similarity between, again, model updates sent by different workers to adapt the learning rate per client based on the update similarity among indicative features in any given iteration. In \cite{bagdasaryan2020backdoor}, they highlight a defense that computes the pairwise cosine similarity between the different workers of the FL process hoping that the attacker’s malicious model update will stand out. While previous defenses compute a vertical examination of the workers between each other, we propose a horizontal examination and combine the use of Euclidean distance and cosine similarity between models sent by the same worker in order to monitor the behaviour of the worker in time and detect a malicious behaviour.

The last defense, attestedFL-3, removes local model updates of \textit{workers} when their performance doesn't improve compared to their own previous performance on the same small validation dataset that the \textit{chief} holds. We use different metrics in order to account for different lines of defense to protect against the attacks. attestedFL-1 observes the Euclidean distance of the local model towards the global model. When the attacker is able to manipulate the magnitude of the weights of the local model updates in order to bypass the first line of defense, attestedFL-2 monitors the cosine similarity by providing a second line of defense. When observing the angular distance of successive local model updates throughout the training of a node, we are able to detect the malicious workers that weren't flagged by attestedFL-1 because the direction of a local model update cannot be manipulated without reducing attack effectiveness.

Algorithm~\ref{alg:REALLY} is implemented at the \textit{chief} node. The attestedFL algorithm uses the three lines of defense attestedFL-1, attestedFL2 and attestedFL-3 for the detection of unreliable \textit{workers} under an untargeted model poisoning attack on a federated learning task. Unlike other works, our defense does not require explicit bounds on the expected number of attackers as it evaluates \textit{workers} individually.

\begin{algorithm}
\newcommand\mycommfont[1]{\footnotesize\ttfamily\textcolor{blue}{#1}}
\SetCommentSty{mycommfont}
\SetAlgoNoLine

\SetKwInput{KwInput}{Input} 
\SetKwInput{KwOutput}{Output}
\SetKwFunction{FMain}{Main}
\SetKwFunction{attestedFLOne}{attestedFL-1}
\SetKwFunction{attestedFLTwo}{attestedFL-2}
\SetKwFunction{attestedFLThree}{attestedFL-3}
\KwInput{Global Model $GM^t$ at iteration \textit{t} that the \textit{chief} sent to the \textit{workers} , Local Model updates $LM_{i,t+1}$ of each \textit{worker} \textit{i}, $H_{i,z}$ a subset \textit{z} of a \textit{worker's} previously uploaded consecutive Local Model update recorded as a pair of $LM$ and $GM$ at that time}
\KwOutput{Reliable Global Model $GM_{t+1}$ at iteration \textit{t}}
\SetKwProg{Fn}{Function}{:}{}
\Fn{\FMain{}}{
\For{Iteration \textit{t}}{
    \For{$\forall$ \textit{workers} \textit{i}}{
        $Reliable = false$\;
        Let $S_t$ be the weight of indicative features at iteration \textit{t}\;
        \If{\attestedFLOne{$H_{i,z}$} = true}{
            \If{\attestedFLTwo{$H_{i,z}$} = true}{
                \If{\attestedFLThree{$H_{i,z}$} = true}{
                    Reliable = true \;
                }
            }
        }
        \If{$Reliable = true$}{
            Let \textit{Real} be the vector containing the index \textit{i} of all reliable \textit{workers} \;
        }
    }
    Federated aggregation of the $LM_{i,t+1}$ \textit{workers} in \textit{Real}\;
   }
}
\vspace{0.4em}
\Fn{\attestedFLOne{$H_{j,z}$}}{
\tcc{See Section~\ref{sec:attestedFL-1}}
\If{$\Delta'_i(t) \leq \mu(t)- 4\sigma(t)$}{
    \Return{true}
}
}
\vspace{0.6em}
\Fn{\attestedFLTwo{$H_{j,z}$}}{
\tcc{See Section~\ref{sec:attestedFL-2}}
\KwInput{$T:$ total training iterations}
$sim_i(t) = \frac{LM_{j}^{t}\{i\} \cdot LM_{j}^{t+1}\{i\}}{\parallel LM_{j}^{t}\{i\} \parallel \parallel LM_{j}^{t+1}\{i\}\parallel}$

$SIM_j(t)=SIM_j(t-1)\cup sim_j(t)$

\If{$\lim_{t \to T} SIM_j(t) \approx 0$}{
    \Return{true}
}
}
\vspace{0.6em}
\Fn{\attestedFLThree{$H_{j,z}$}}{
\tcc{See Section~\ref{sec:attestedFL-3}}
\KwInput{$T:$ total training iterations}
$E_j^{t} - E_j^{t-1}= E_j^{t}(LM_j^{t}) - E_j^{t-1}(LM_j^{t-1})$

$E_j(t) = E_j(t-1)\cup {\{E_j^{t} - E_j^{t-1}\}}$

\If{$\lim_{t \to T} E_j(t) \approx 0$}{
    \Return{true}
}
}
\vspace{0.6em}

\caption{attestedFL algorithm using attestedFL-1, attestedFL-2 and attestedFL-3 to remove local model updates of unreliable \textit{workers}.}
\label{alg:REALLY}
\end{algorithm}


\subsection{attestedFL-1}
\label{sec:attestedFL-1}

The insight in this work is that when a shared model is under an untargeted attack, monitoring the \textit{workers} history of updates will expose an attacker's malicious behavior in time and show that it is not training, not advancing towards a goal.  

\subsubsection{Convergence of the models}

attestedFL-1 proposes to monitor the history of updated local models in comparison to the global model as training advances. Precisely, we aim at observing the convergence of the Euclidean distance of the local model towards the global model. Let us define the Euclidean distance of the local model of a \textit{worker} $j$ and the global model at the iteration $t$ as in Equation~\ref{eq:conditioneq2}
\begin{equation}
\label{eq:conditioneq2}
\Delta_{j}(t) = \parallel LM_{j}^{t+1} - GM^t \parallel
\end{equation}


Let $N$ be the set of all \textit{worker} nodes, such that $|N|=n$ and let $N_b\subset N$ be the subset of $b$ benign \textit{workers}. For benign \textit{workers}, as $t$ increases, $\Delta(t)$ tends to a stable value, preferably to a very small value as training advances, i.e., if $T = \{1,2,\ldots,t_f\}$ is the total number of training steps, $t_c \in T$ is a given step of the training and $j \in N_b$ is a \textit{worker}, then $\Delta$ behaves as in Equation ~\ref{eq:delta_bening}.
\begin{equation}
\label{eq:delta_bening}
\lim_{t_c \to\ t_f} \Delta_{j}(t_c) \sim 0
\end{equation}
It is observed that $t_c$ is the first iteration step when $\Delta(t)$ is computed by the \textit{chief}. Before iteration $t_c$, there is a warm-up period for the \textit{workers} to start converging towards the global model. This is to account for the fact that before $t_c$ the local models of all \textit{workers} may be too far from the global model.

Let $j$ be a malicious \textit{worker}, i.e., $j\in N \setminus N_b$. In targeted attacks, since the attacker tries to deviate from the global model with a fixed Malicious Model $MM^t$, $\Delta$ is expected to increase as training advances, i.e., $\Delta$ behaves as in Equation ~\ref{eq:delta_malicious}.
\begin{equation}
\label{eq:delta_malicious}
\lim_{t_c \to\ t_f} \Delta_{j}(t_c) \sim x
\end{equation}
where, $x \in \mathbb{R}$ and is sufficiently large (but finite).

In untargeted attacks, $\Delta$ is randomly distributed. An attacker targeting availability will in expectation over the entire training process not decrease in the difference of its residual in comparison to the global model. The difference will be randomly distributed as an attestation that the \textit{worker} is not training, hence the $\lim_{t_c \to\ t_f} \Delta_{j}(t_c)$ does not exist.









\subsubsection{Convergence Speed}

In addition to the convergence of the model, we wish to capture the rate of decrease to assess convergence speed. The rate of decrease captures an attack where even if $\Delta_j$ tends to zero the total effect of the adversarial is significant. For example, if the convergence of the model is a harmonic sequence such that $LM_{j}^{t+1} - GM^t = 1/t$, then the total effect of this attack is still non-zero. By only relying on $\Delta_j$, we could not capture behaviors such as the harmonic behavior. 

To detect this type of behavior, we introduce the convergence speed. Precisely, the average rate of decrease corresponds to the changes in $\Delta_j$, while the model is training. Knowing that at the beginning of training, the weights of local models tend to considerably diverge from those of the global model, we propose to attribute more and more importance to the $\Delta$ values as the training advances. 
Hence $t_c \leq t$ the weighted-average rate of decrease is defined as follows:
\begin{equation}
\Delta'_j(t) = \frac{\partial \Delta_j}{\partial t} = \frac{\sum\limits_{t_c}^{t} \left(1-\exp^{-\frac{t}{c}\left(\Delta_{j}(t+1)-\Delta_{j}(t)\right)}\right)}{c}
\end{equation}



In fact, $\Delta'_j(t)$ represents the sum of the weighted-average of the difference in convergence for the $t - t_c = c$ previous iterations of the training, i.e., $\Delta'_j(t)$ measures how fast the local model of $j$ is converging to the global model. The weights in the index around the current iteration increase so as to better capture training behavior. In fact, the closer $t$ is to the current training step, the more the contribution to the average rate of decrease. This represents a better assessment of the cumulative distribution.


At each iteration $t$, we then compute the mean, $\mu(t)$, of $\Delta_j(t)$ for all $j\in N$ and we also compute its standard deviation, $\sigma(t)$. Inspired from the assumptions in \cite{fang2020local} about each \textit{i}th parameter of the local model update of the benign \textit{worker} being a sample from a Gaussian distribution with mean $\mu_i$ and standard deviation $\sigma_i$, Fang et al. estimate that the global model weights are $w_{max,i} < \mu_i + 4\sigma_i < w_{min,i}$ with large probabilities. We use the same bounds as reference to exclude nodes having rate of convergence that fall under $\mu(t)- 4\sigma(t)$. Hence in Algorithm~\ref{alg:REALLY}, the conditions implemented for attestedFL-1, for the node $j$ at iteration $t$ are as follows:

\begin{equation}
    \text{attestedFL-1} = 
\begin{cases}
    \text{accept }j & \text{if } \Delta'_j(t) \leq \mu(t)- 4\sigma(t)\\
    \text{reject }j & \text{otherwise}
\end{cases}
\end{equation}

It is expected that $\Delta'_{j}(t) < \Delta'_{k}(t)$ if $j \in N_b$ and $k \in N\setminus N_b$, i.e., the convergence speed of benign nodes will always be faster than the convergence speed of malicious nodes performing an untargeted attack. 



\begin{itemize}

\item Within certain bounds over time, the effect of certain behavior such as the harmonic can be captured by the rate of decrease. As the harmonic effect becomes more visible, convergence won't be quick, decreasing the rate, the node becomes more and more an outlier and gets rejected by attestedFL-1.


\end{itemize}

attestedFL-1 would force attack weights to show a trend in learning while trying to maximally mislead the estimated model. This adds an additional challenge for the attacker to choose from iteration to iteration weights inline with the training process expected at the local node, thus becoming less focused towards the poisoning objective. We note that if the attacker ``always'' agrees with the global model, this behavior, although not desirable since the local data is not useful, does not make the global model converge on the FL task. In other words, this behavior will not lead to a successful attack. We do not detect this type of attacker as here we only address the model-poisoning attacks that target the model availability by disrupting the training on the FL task. 

\subsection{attestedFL-2}
\label{sec:attestedFL-2}
In this section, we propose attestedFL-2, a technique that also aims at assessing the reliability of the \textit{worker} during training. The approach is to discard unreliable nodes from the aggregation process and maintain \textit{workers} that seem to be training. This will devalue contributions from \textit{workers} that are not training and will enable the FL process to converge faster. 

While attestedFL-1 and attestedFL-2 try to detect a \textit{worker} that is failing to train towards a consistent goal, they evaluate different behaviors of a malicious \textit{worker} and use different metrics and strategies to do so. attestedFL-2 accounts for conditions under which attestedFL-1 would fail to detect the attack. In attestedFL-2, we measure the cosine similarity of successive local model updates, and observe how the angular distance is behaving throughout the training of a node. Although, the model updates from different nodes are almost orthogonal to each other with very low variance, this is not the case for model updates provided by the same node as training advances. Iteration after iteration the same node is sending local model updates that are more correlated between each other than with model updates of other nodes. A node that is not exhibiting some type of correlation over time between its local model updates may be considered as unreliable. 

Particularly, we use cosine similarity on the indicative features. This means that we consider only the values in the output layers of the model because they map directly to output probabilities. Let $sim_j(t)$ be the cosine similarity on the indicative features of successive local models of \textit{worker} $j$ at iteration $t$, it is defined as follows: 
\begin{equation}					
sim_j(t) = \frac{LM_{j}^{t}\{i\} \cdot LM_{j}^{t+1}\{i\}}{\parallel LM_{j}^{t}\{i\} \parallel \parallel LM_{j}^{t+1}\{i\}\parallel},
\end{equation}
where $LM_{j}^{t}\{i\}$ are the $i$ indicative features of $LM_{j}^{t}$ 

Every iteration, attestedFL-2 extracts a subset of a \textit{worker's} previously uploaded consecutive local model updates and computes the cosine similarity between successive updates, i.e., at iteration $t$ for all $j\in N$ the \textit{chief} computes the set $SIM_j(t) = \{sim_j(t-c), sim_j(t-(c+1)), \ldots, sim_j(t-1), sim_j(t)\}$.

Our defense looks at a pattern in $SIM_j(t)$ and not just the local successive updates because the variance in the updates of two consecutive iterations causes the cosine similarities at each iteration to be an inaccurate approximation of a \textit{worker's} malicious likelihood. In Algorithm~\ref{alg:REALLY}, the conditions for attestedFL-2 are as follows:

\begin{itemize}
\item For benign nodes that are training, the cosine similarity metric attests to the similarity of the successive $LM_j$ updates as training advances. In this case, the elements of the set $SIM_j(t)$ increase their cosine similarity over the iterations. 

\item Under an untargeted attack, the cosine similarity score between local model updates of a single \textit{worker} present lower and lower values as training advances. The elements of the set $SIM_j(t)$ decrease their cosine similarity at each iteration, a progression is not observable in this case.

\end{itemize}

\subsection{attestedFL-3}
\label{sec:attestedFL-3}
Federated learning explicitly assumes that participants’ local training datasets are relatively small and drawn from different distributions. Since non-i.i.d. data are used in FL, each local model may be quite different from the global model. Thus, there are significant differences between the weights of individual models. FL takes advantage of the diversity of participants with non-i.i.d. training data, including unusual or low-quality local data, and by design, the \textit{chief} should accept even local models that have low accuracy and significantly diverge from the current global model. 

In attestedFL-3, to assess if the \textit{worker} is reliable, assuming that the \textit{chief} has a small validation dataset, the \textit{chief} can test to see how the local model of a \textit{worker} predicts on the validation dataset. From local model update to another sent by the \textit{worker}, the \textit{chief} tests for the same sample set the \textit{worker}'s performance. This will highlight if the node appears to be learning. The intuition is that on a same sample, we evaluate the performance over a set of iterations within a time window to see how they individually perform compared to their own previous performance on that validation dataset and not compared to other nodes because every node is learning differently.

Precisely regarding the small validation dataset, Pan et al. \cite{pan2020justinian} introduced the concept of \textit{quasi-validation set}, a small dataset that consists of data samples from similar data domains. Inspired from their work, we use a \textit{quasi-validation set} that represent a collection of data samples that follows a similar, but not necessarily identical distribution as the true sample distribution. In practice, if a gold-labeled validation set (i.e., a set of samples from the true sample distribution) is available during the learning process, it can be used as a \textit{quasi-validation set}. Otherwise, the \textit{chief} can randomly collect a small number of validation data samples from similar data domains to form the \textit{quasi-validation set}.

We compute the impact of each local model on the error rate for the \textit{quasi-validation set} and remove the local models that have large negative impact on the error rate. Specifically, at iteration $t$, for each \textit{worker} $j$, we compute the error, $E_j^{t+1}$, of the current local model $LM_j^{t+1}$ on the \textit{quasi-validation set} and for the same \textit{worker} we compute the error of the local model at the previous iteration, i.e. $E_j^{t}$.
Hence the error rate impact of a local model is defined as follows: 
\begin{equation}
E_j^{t+1} - E_j^{t}= E_j^{t+1}(LM_j^{t+1}) - E_j^{t}(LM_j^{t})
\end{equation}

The conditions implemented in Algorithm~\ref{alg:REALLY} for attestedFL-3 are as follows:

\begin{itemize}
\item If the node is training well, its performance on the \textit{quasi-validation set} will improve, i.e., $E_j^{t+1} - E_j^{t}$ will be smaller over time and the node will be considered reliable. 

\item If $E_j^{t+1} - E_j^{t}$ increase over time, then the node is considered unreliable. This is due to the fact that attackers are not concerned in training locally their model but rather fabricating update values.
\end{itemize}

It is important to note that a \textit{worker} is considered unreliable only when its performance on the \textit{quasi-validation set} decreases. If the performance stays constant or doesn't improve, we do not consider the \textit{worker} as malicious. This prevents a \textit{worker} from being wrongly disqualified in the advent of the \textit{chief's} set being not diverse.

\section{Evaluation}
\label{sec:evaluation}
To demonstrate the untargeted attack, we implement three different federated learning settings each using a different dataset. The first two settings are for different benchmark systems for text and image classification. The text use case consists in training a fully connected neural network for the hand-written digital classification task on the MNIST dataset. This public dataset contains 60,000 28×28 images of 10 digits for training and 10,000 for testing. The model consists of 784 inputs, 10 outputs with soft-max activation and one hidden layer with 30 ReLu units. The dimension of parameters is 25,450. The image use case is training a ResNet-18 model for the image classification on the CIFAR-10 dataset. This dataset contains 60,000 28×28×3 images of 10 classes of objects for training and 10,000 for testing. The standard model ResNet-18 has 18 end-to-end layers and 11,173,962 learnable parameters in total. In all use cases, each \textit{worker} shares a copy of the training set. We simulate the federated learning setting by sequential computation of gradients on randomly sampled mini-batches. 

The third FL setting aims at evaluating a more realistic deployment and consists of training a Convolutional Neural Network (CNN) classifier for mobility mode inference as in \cite{Lopez2019BCFL}. The CNN architecture used in the work is that of~\cite{yazdizadeh2019ensemble} and is summarized as follows:
\begin{enumerate}
    \item $5$ convolutional layers, each one followed by a max-polling layer
    \item Convolution kernels of size $8$ and a max-pooling operation of size $2$
    \item Number of Kernels equal to $96$, $256$, $384$, $384$ and $256$ for five convolutional layers
\end{enumerate}


A subset of open access trajectory data gathered by the city of Montr\'eal using a smartphone app, \textit{MTL Trajet}~\cite{yazdizadeh2019ensemble} is used to train the federated CNN for transportation mode inference. The \textit{MTL Trajet} dataset contains the coordinates collected by the GPS sensor of smartphones and mode prompt data collected by users after they finished a trip. The training dataset consists of trip information from $10$ users collected during the fall of 2016. The trip data of each traveler is assigned to each \emph{worker}, so the final training dataset consists of $321$ ($80\%$) trip segments and $87$ trip segments in the test set ($20\%$) comprise the training and test datasets. To evaluate the performance of our defense when the FL process is under attack, we study the accuracy the model is attaining during training. We refer to the different FL settings as MNIST, CIFAR and MTL Trajet.

\subsection{Experimental setup}
The setup consists in an \emph{Amazon EC2 t3.2xlarge Virtual Machine} used to run the nodes who are participating in the FL process, where one node is designated as \textit{\textit{chief}} while the others are \textit{\textit{workers}}.

As a base case scenario, we run the FL model under no attack and the accuracy at each epoch is reported. In the baseline scenario, the FL process is expected to end with a trained model at a good accuracy. We present results for various attack strategies of continuous untargeted attacks, adaptive attacks, single attacker, two attackers colluding and attacks at different stages of convergence. Performing attacks at different stages of convergence enables us to better assess the performance of the defense mechanism.

\subsection{Experimental Results}


\subsubsection{Defense against continuous untargeted attacks}

\begin{figure}[!h]
    \centering
    \includegraphics[width=1\columnwidth]{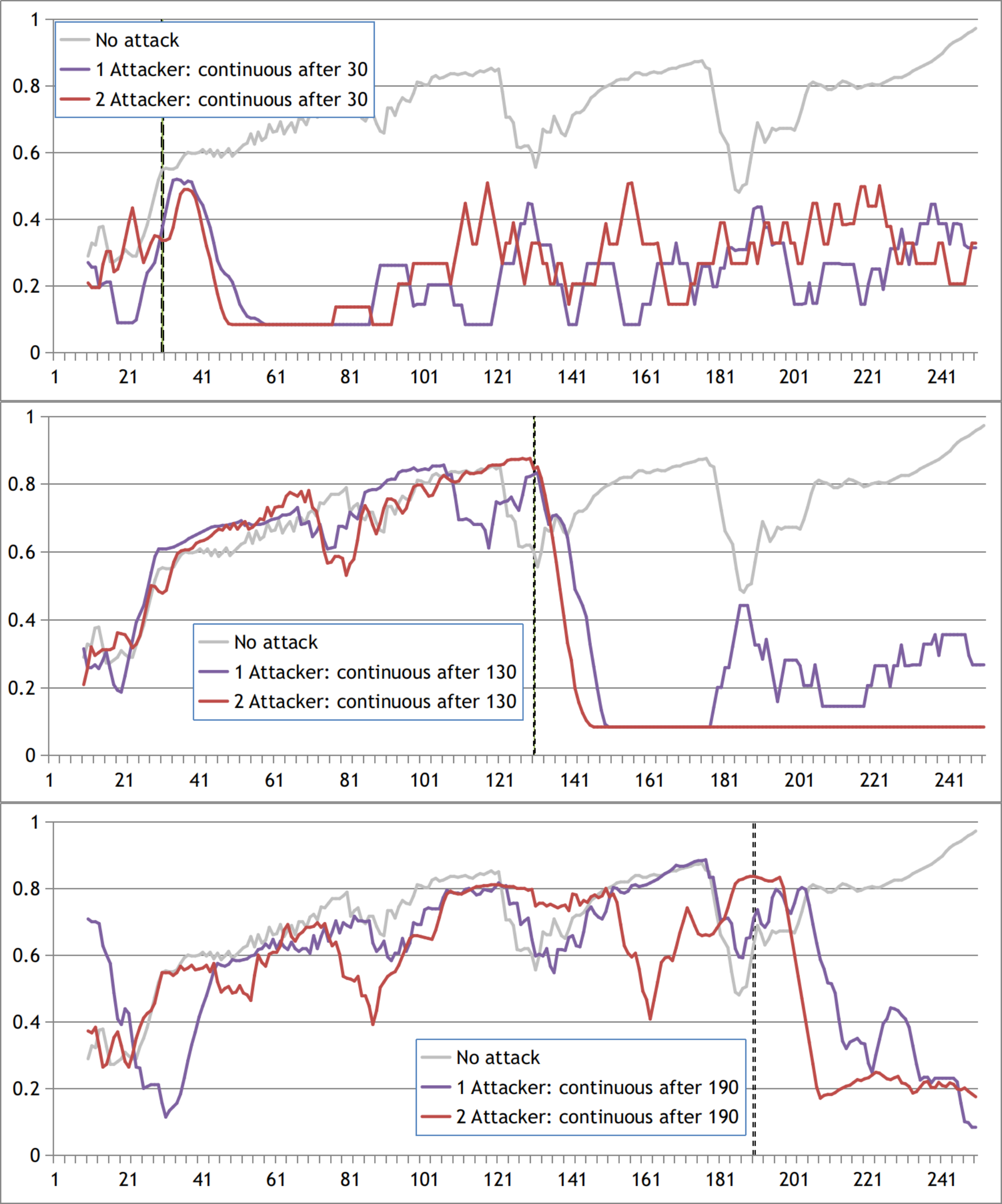}
    \caption{Accuracy of the MTL Trajet CNN model under a continuous untargeted attack after the epoch $30$, $130$ and $190$ by one and two \textit{workers} for every training iteration.}
    \label{fig:untargeted_continuous_attack}
\end{figure}

We first test the impact of our defense in the CNN FL setting on MTL Trajet dataset. Continuous untargeted attacks are implemented after epochs $30$, $130$ and $190$ by one and then by two \textit{workers}. The model accuracy is impacted when one or two \textit{workers} send continuous untargeted attacks at epoch $30$, $130$ and $190$, respectively (Figure~\ref{fig:untargeted_continuous_attack}). We observe that the accuracy in these three tests drops, but shows recovery after some epochs. The model tends to recover better from attacks at the early stages of training. This behavior is expected since attacks injected, as the global model is converging, tend to stay for a longer period of time. 


\begin{figure}
    \centering
    \includegraphics[width=1\columnwidth]{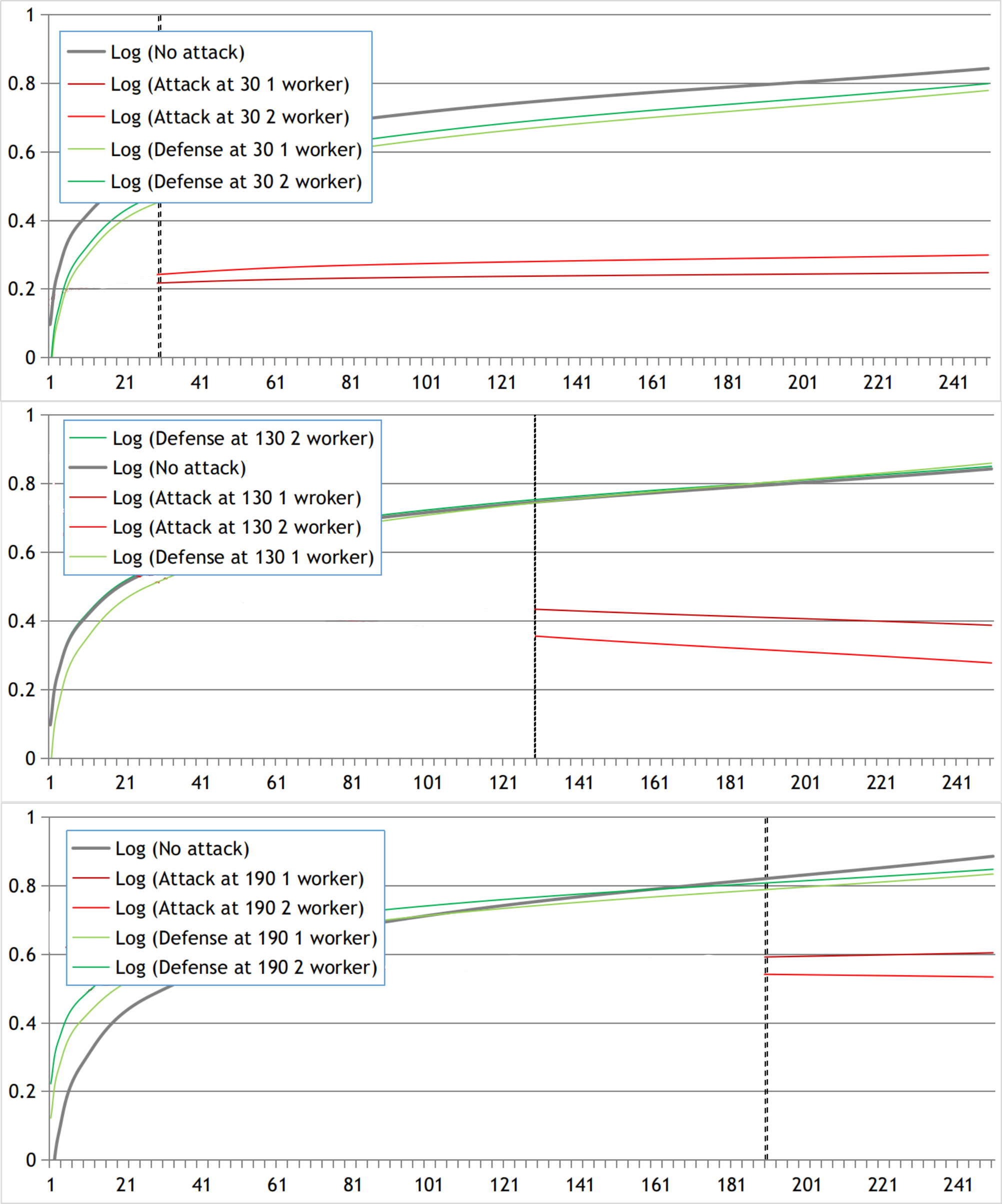}
    \caption{Convergence of the MTL Trajet CNN model during continuous untargeted attacks at EPOCHs 30, 130 and 190 by one and two \textit{workers} in comparison to the convergence with attestedFL-1 for every training iteration (logarithmic trend fit).}
    \label{fig:untargeted_continuous_convergence_attestedFL1}
\end{figure}

To defend against this scenario of continuous untargeted attacks, we implement the attestedFL algorithm. At each epoch, attestedFL tests if nodes are sending reliable weights by examining their past training behavior. To better assess the impact of attestedFL-1, we present in Figure~\ref{fig:untargeted_continuous_convergence_attestedFL1} the convergence of the attacks in comparison to that of the defense. We notice that the accuracy was decreasing under attack, but increased at an average of $41\%$ by attestedFL-1. While the attack tries to decrease convergence, attestedFL-1 increases the accuracy, the model can achieve under this adversarial setting. The defense shows promising results for attacks at any stage of convergence. This is due to the fact that the defense looks at a pattern in the training and eliminates misbehaving nodes. 

\begin{figure}
    \centering
    \includegraphics[width=1\columnwidth]{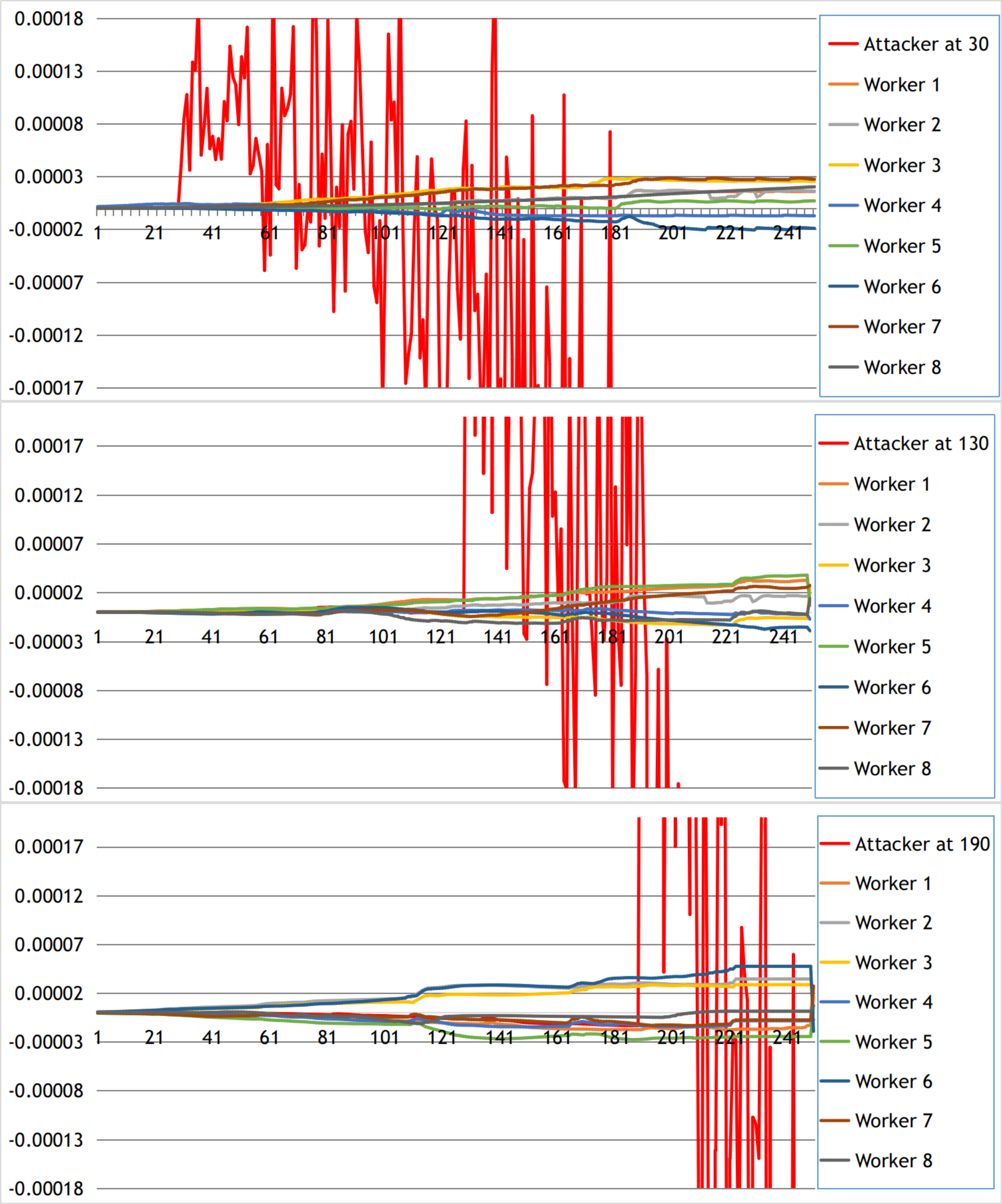}
    \caption{Variation of $\Delta'_j(t)$ of the benign \textit{workers} and the attackers during continuous untargeted attacks on the MTL Trajet CNN model at epoch 30, 130 and 190 done by one and two \textit{workers} for every training iteration.}
    \label{fig:untargeted_continuous_deltaandspeed}
\end{figure}

To better observe the pattern in the training, we present in Figure~\ref{fig:untargeted_continuous_deltaandspeed} the variation of $\Delta'_j(t)$ as $t$ increases.
We notice that benign nodes are faster at converging than the malicious nodes performing an untargeted attack, in fact the speed of convergence of malicious nodes shows variations but with a trend of decelerating its convergence.

\begin{figure}
    \centering
    \includegraphics[width=1\columnwidth]{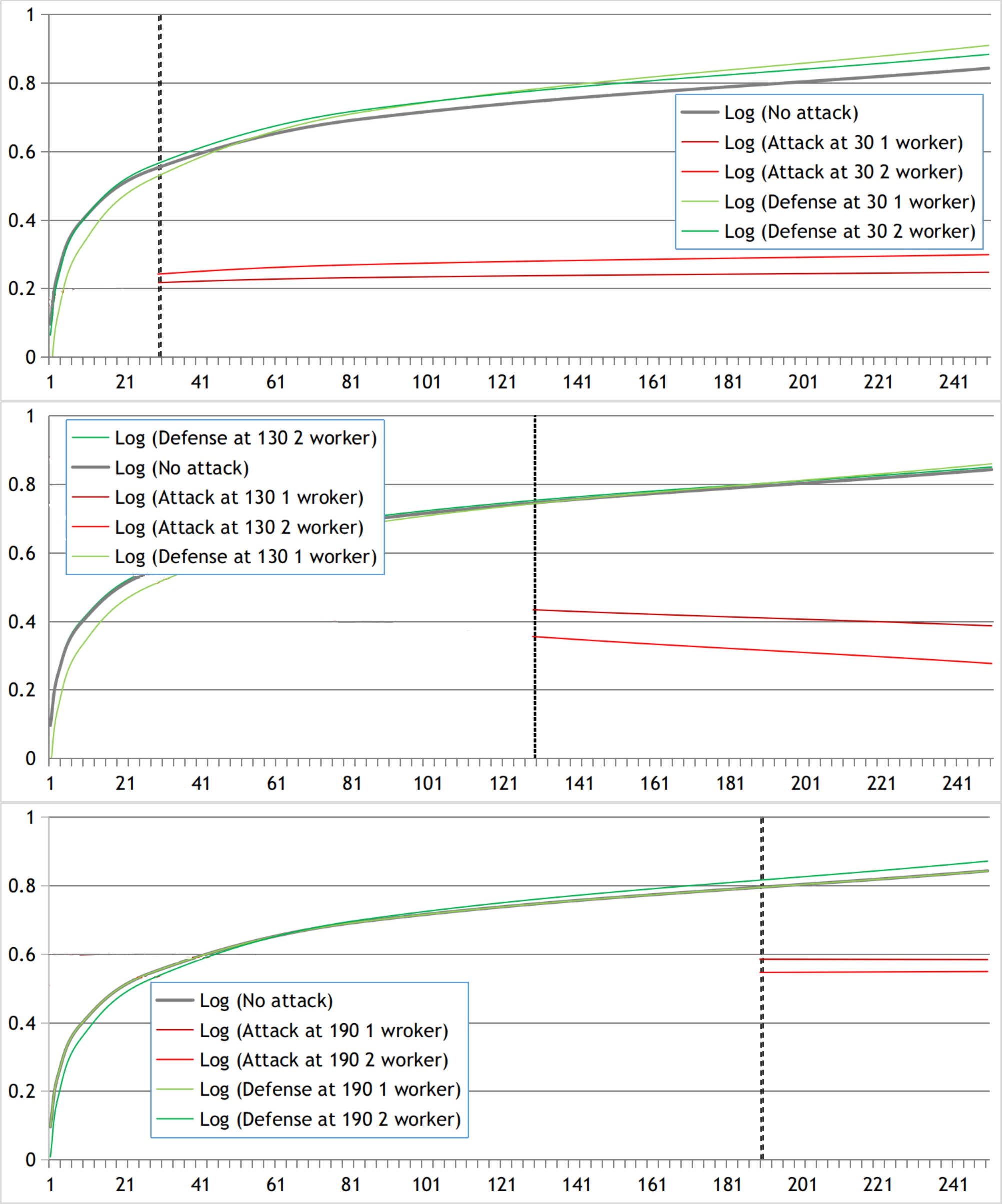}
    \caption{Convergence of the MTL Trajet CNN model during continuous untargeted attacks at epoch 30, 130 and 190 by one and two \textit{workers} in comparison to the convergence of attestedFL-2 at every training iteration (logarithmic trend fit).}
    \label{fig:untargeted_continuous_convergence_attestedFL2}
\end{figure}

To assess the impact of attestedFL-2, we present in Figure~\ref{fig:untargeted_continuous_convergence_attestedFL2} the convergence of the attacks in comparison to that of the defense. We notice that the accuracy was decreasing under attack, but increased at an average of $45\%$ by the defense. attestedFL-2 increases the accuracy the model reaches under this adversarial setting.

\begin{figure}
    \centering
    \includegraphics[width=1\columnwidth]{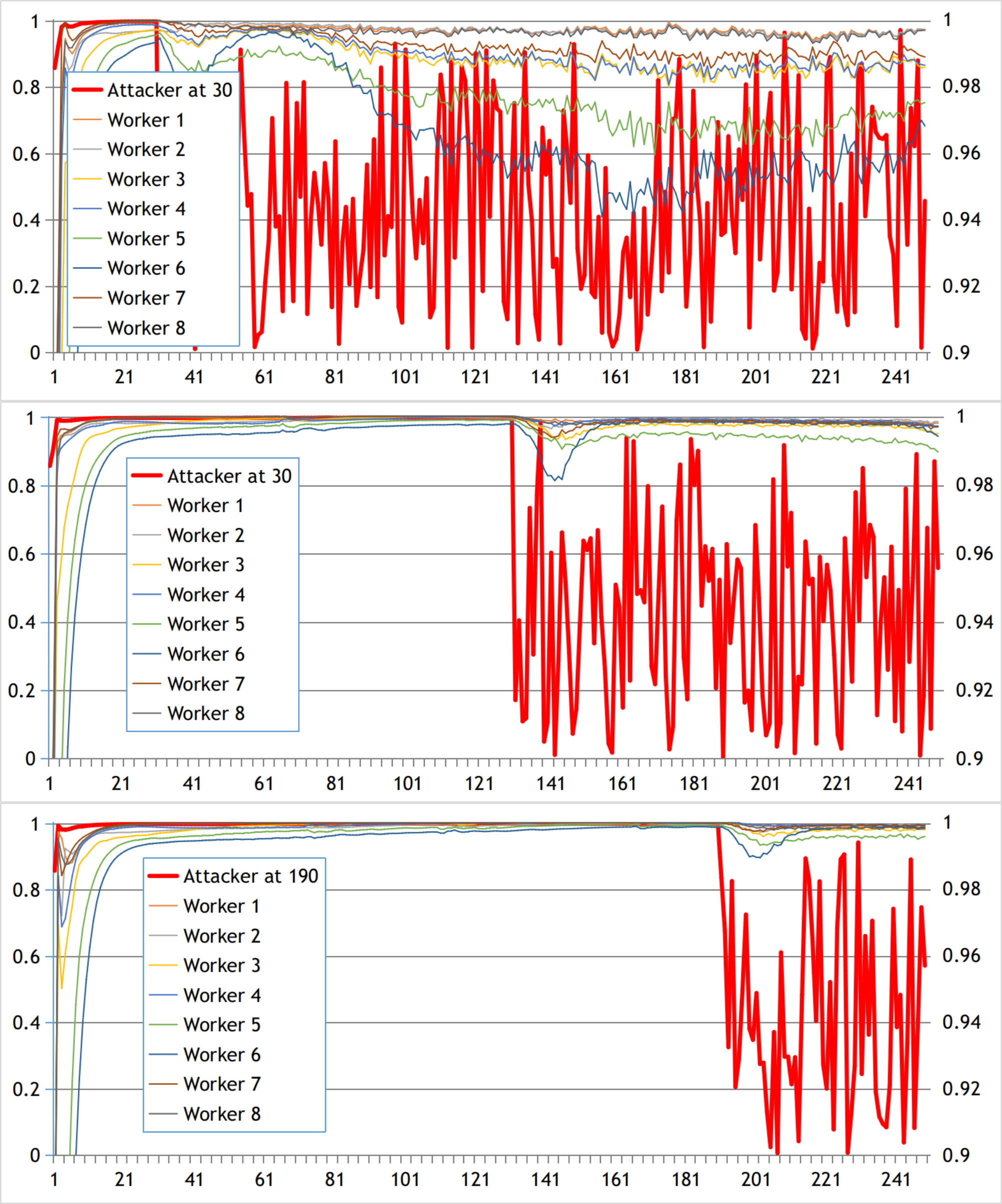}
    \caption{Cosine similarity of one attacker and eight \textit{workers} in an untargeted continuous attack scenario at epoch 30, 130 and 190 for every training iteration.}
    \label{fig:untargeted_Continuous_cosine}
\end{figure}

Figure~\ref{fig:untargeted_Continuous_cosine} shows the evolution of the cosine similarity during training of the different nodes of the system in this scenario. We see how the attacker is not training, thus its cosine similarity is sparsely distributed between $0$ and $1$. As opposed to benign \textit{workers}, where the cosine similarity between consecutive local model updates is distributed between $0.95$ and $1$ as a good indication that the \textit{worker} is training. 

\begin{figure}
    \centering
    \includegraphics[width=1\columnwidth]{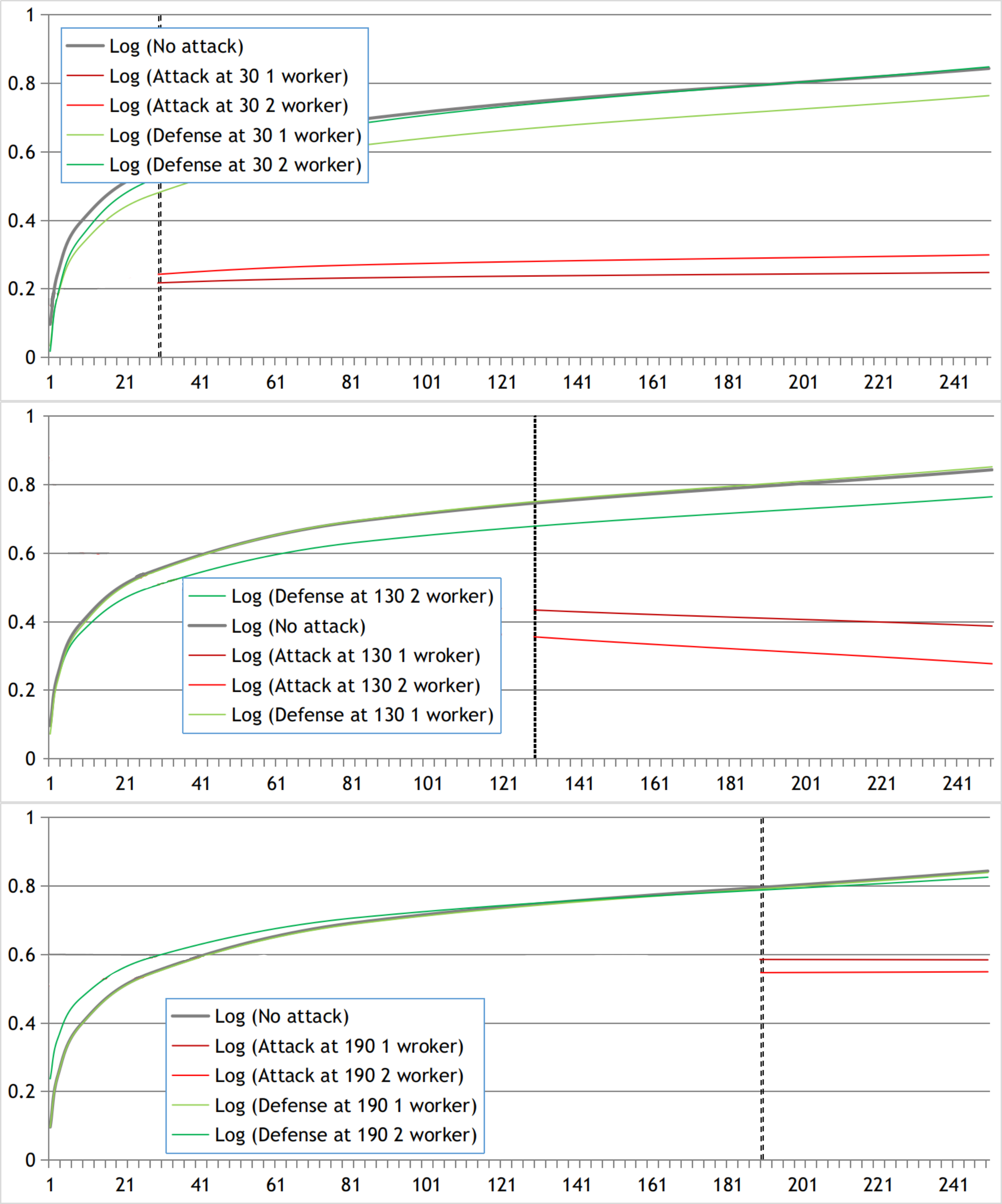}
    \caption{Convergence of the MTL Trajet CNN model during continuous untargeted attacks at epoch 30, 130 and 190 by one and two \textit{workers} in comparison to the convergence of attestedFL-3 for every training iteration (logarithmic trend fit).}
    \label{fig:untargeted_continuous_convergence_attestedFL3}
\end{figure}

\begin{figure}
    \centering
    \includegraphics[width=1\columnwidth]{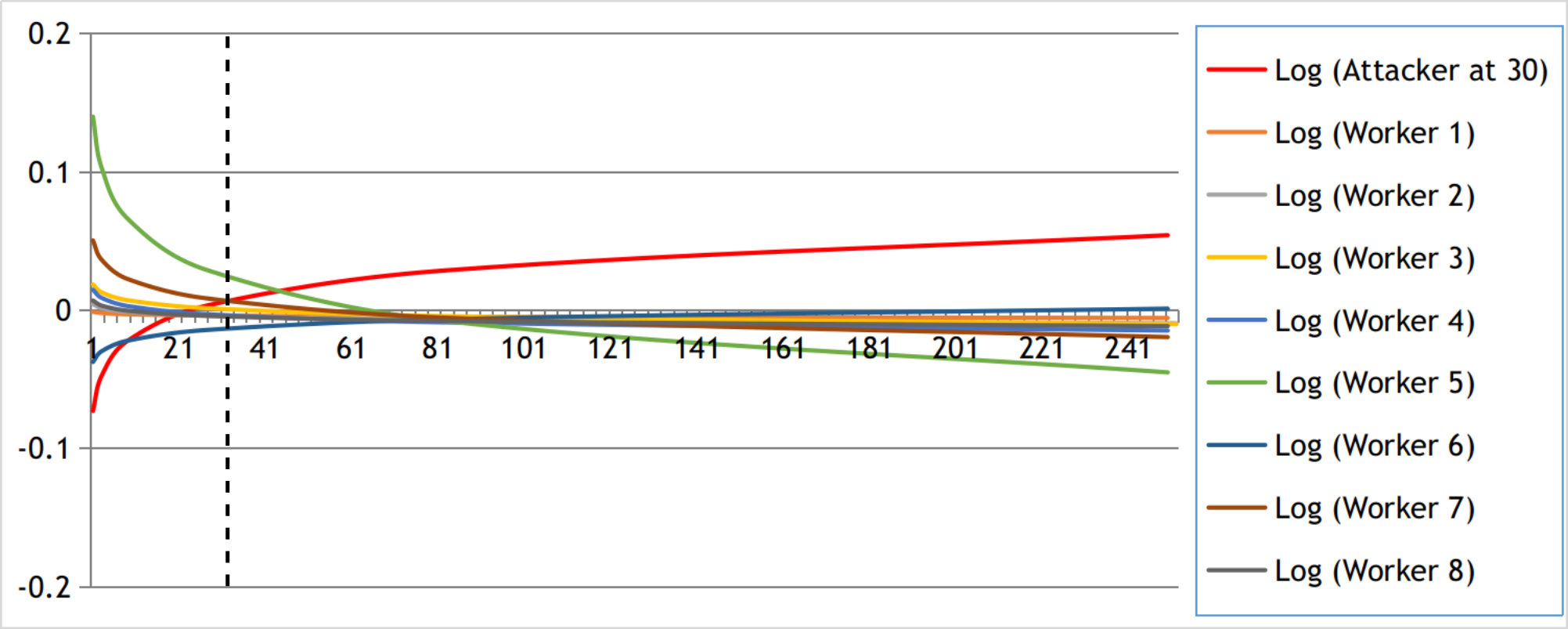}
    \caption{Impact of each local model on the error rate for benign and malicious \textit{workers} during continuous untargeted attacks on the MTL Trajet CNN model at epoch 30, 130 and 190 done by one and two \textit{workers} for every training iteration (logarithmic trend fit).}
    \label{fig:untargeted_Continuous_error}
\end{figure}

To assess the impact of attestedFL-3, we present in Figure~\ref{fig:untargeted_continuous_convergence_attestedFL3} the convergence of the attacks in comparison to that of the defense. We notice that the accuracy was decreasing under attack, but increased at an average of $43\%$ by the defense.
Figure~\ref{fig:untargeted_Continuous_error} shows the evolution of the error rate between consecutive iterations of benign and malicious nodes. As training advances, a larger error rate impact between consecutive iterations indicates that the \textit{worker} is not training and should be removed. If the node is training well, its performance on the \textit{quasi-validation set} will improve, i.e., $E_j^{t+1} - E_j^{t}$ will be smaller over time and the node will be considered reliable.

\begin{figure}
    \centering
    \includegraphics[width=1\columnwidth]{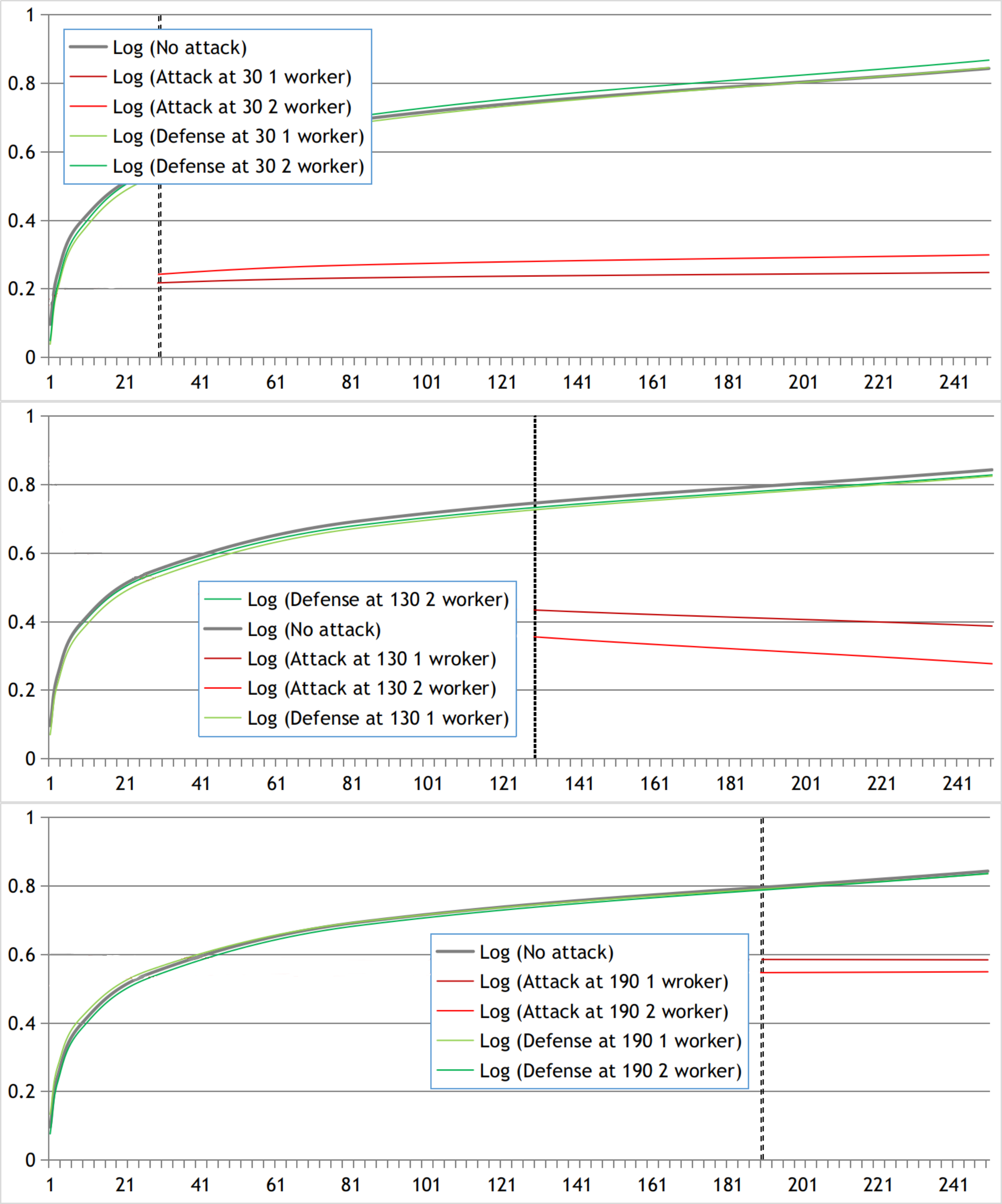}
    \caption{Convergence of the MTL Trajet CNN model during continuous untargeted attacks at epoch 30, 130 and 190 by one and two \textit{workers} in comparison to the convergence under the three lines of defense (attestedFL-1, attestedFL-2, attestedFL-3) for every training iteration (logarithmic trend fit).}
    \label{fig:untargeted_continuous_convergence_attestedFL123}
\end{figure}

We also apply all three lines of defense attestedFL-1, attestedFL-2 and attestedFL-3 to protect against the untargeted poison attack and present in Figure~\ref{fig:untargeted_continuous_convergence_attestedFL123} the convergence of the attacks on the MTL Trajet CNN model in comparison to that of the defense. We noticed that the accuracy was decreasing under attack, but increased at an average of $50\%$ by the defense. 

\subsubsection{Impact of the defense in different FL settings}

In this section, we present the results of attestedFL under different FL settings in order to demonstrate that our proposed defense is effective across many datasets. We vary the number of rounds and show the effect of our defense for different benchmark systems for text and image classification, MNIST and CIFAR. 

\begin{figure}
    \centering
    \includegraphics[width=1\columnwidth]{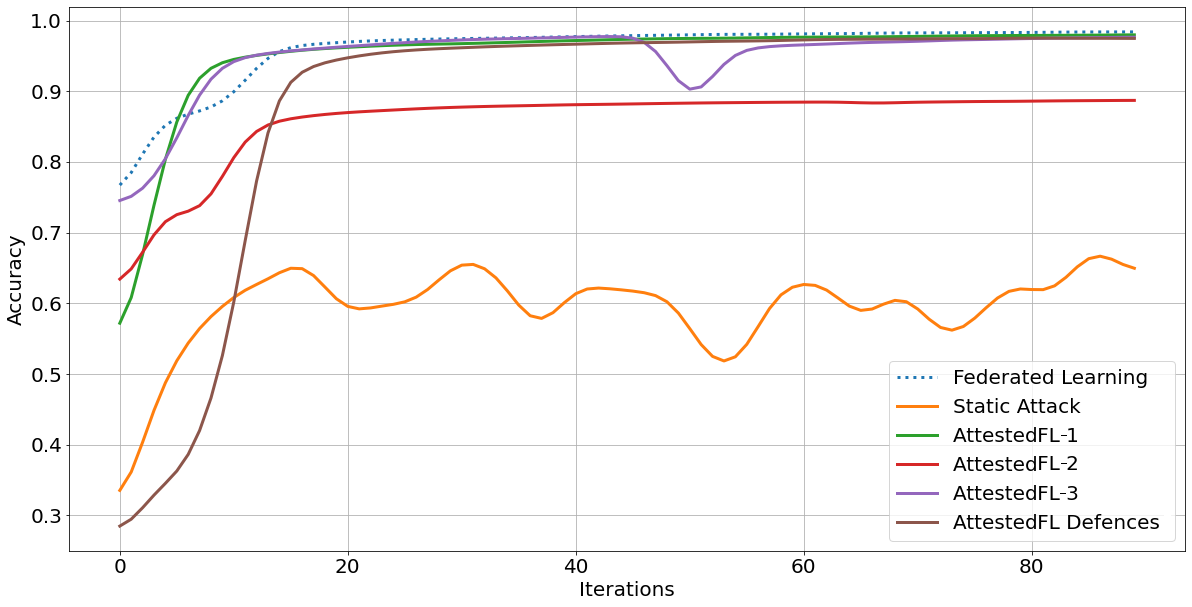}
    \caption{Effect of the three lines of defense (attestedFL-1, attestedFL-2, attestedFL-3) against an attack on the MNIST model.}
    \label{fig:MNIST_attestedFL123}
\end{figure}

Figure~\ref{fig:MNIST_attestedFL123} shows how the model accuracy is impacted by the untargeted static attack on MNIST compared to the blue baseline curve of the federated learning process under no attack. In this scenario, when activated individually, we notice that attestedFL1, attestedFL2 and attestedFL3 are able to protect against the attack by improving the accuracy of the model by an average of $98.5\%$, $75.7\%$ and $97.1\%$ respectively compared to the attack scenario. Against sophisticated attacks, we can activate all three lines of defense at the same time. In this scenario, we started all attestedFL defenses only after 10 iterations of training and the results for MNIST show that they provide the best protection in terms of convergence of the model.

\begin{figure}
    \centering
    \includegraphics[width=1\columnwidth]{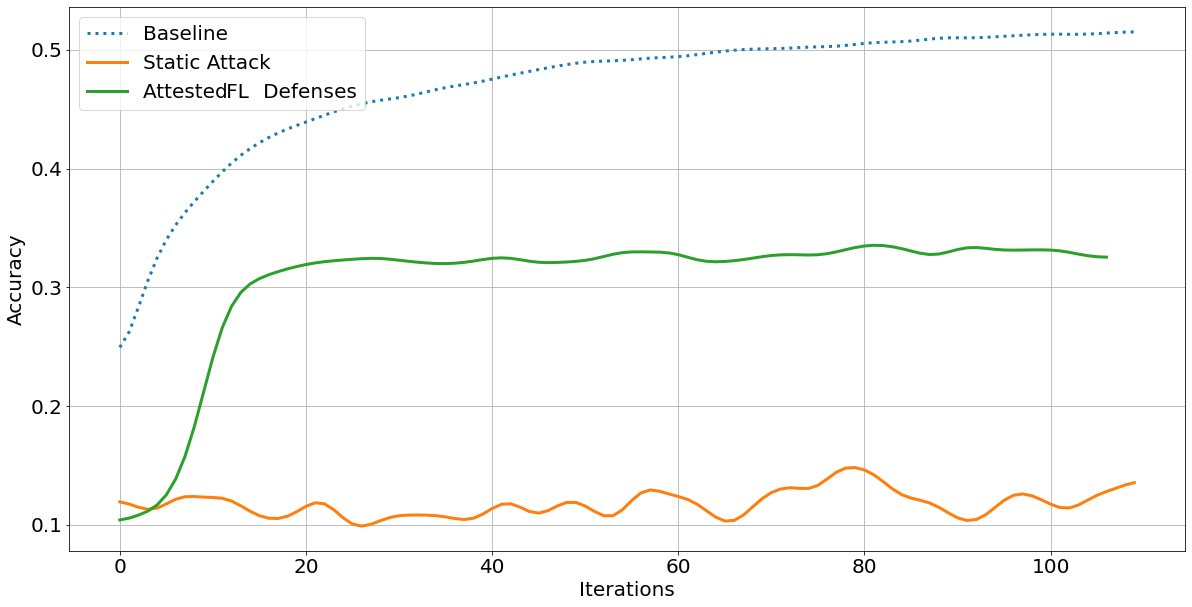}
    \caption{Effect of the three lines of defense (attestedFL-1, attestedFL-2, attestedFL-3) against an attack on the CIFAR model.}
    \label{fig:CIFAR_attestedFL123}
\end{figure}

Figure~\ref{fig:CIFAR_attestedFL123} shows the accuracy of the model at every iteration for CIFAR. We notice the impact of the attack compared to the baseline and how the three lines of defense were able to increase the accuracy. In this scenario, the accuracy increased at an average of $53.4\%$ when all defenses were activated at the same time.

\subsubsection{Adaptive attacks}
We test the effectiveness of attestedFL under different attacking patterns that create difficulties in our defense. In fact, an adversary's knowledge of the different layers of attestedFL would force attack weights not only to be chosen to maximally mislead the estimated model, but also to show a trend in learning. Moreover, iteration after iteration the same node must send local model updates that are more correlated between each other to show some type of correlation over time between its local model updates.

First, we investigate a static attack where a fixed number of \textit{worker} are compromised and perform continuous attacks during the entire learning process. Another attack pattern consists of a fixed number of \textit{workers} pretending to be benign in the first fixed number of rounds and start the attack afterwards. We call this attack the pretence attack. Finally, we proposed an attack pattern where each compromised \textit{worker} is assigned with its role by the adversary. During the learning process, the \textit{worker} changes its role with a certain probability. In this case, the malicious node maintains two concurrent local models, one malicious, the other benign. We call this attack a randomized attack where the adversary could choose to not intercept and replace a node’s real updates to avoid detection. 

\begin{figure}
    \centering
    \includegraphics[width=1\columnwidth]{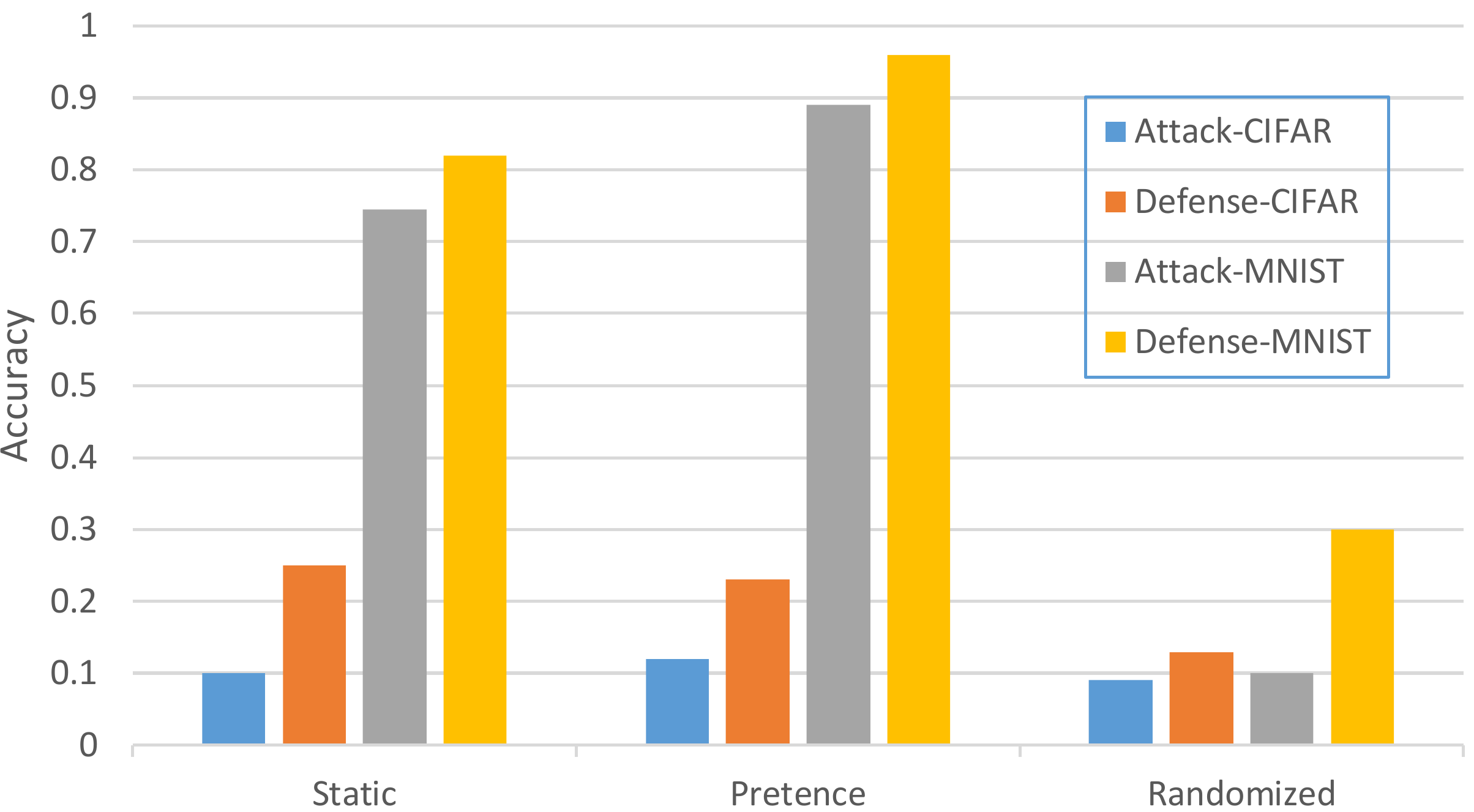}
    \caption{Accuracy of the global model under a static, pretence and randomized attack for MNIST and CIFAR defended with attestedFL-3.}
    \label{fig:patterns}
\end{figure}

Figure~\ref{fig:patterns} shows the accuracy of the global model when under a static, pretence and a randomized attack for MNIST and CIFAR defended with attestedFL-3. The results show that under any attacking pattern, the defense increased accuracy because it is adding an additional  challenge for the  attacker layer after layer. attestedFL proved to be effective because it was able to increase the accuracy between 7\% and 67\% for MNIST and CIFAR with only one of its layers of defense. In fact, the attacker must choose from iteration to iteration weights inline with the training process expected at the local node, thus becoming less focused towards the poisoning objective.

Finally, we implement a targeted attack and evaluated the performance of attestedFL in this scenario. Under targeted attacks, the goal of the adversary is to ensure that the learned model behaves differently on certain targeted sub-tasks while maintaining good overall performance on the primary task. Hence, we considered the targeted model update poisoning attack as in \cite{bagdasaryan2020backdoor}. The attack tries to guarantee the replacement of the global model by a malicious model. In this case, attackers may collude and together, every epoch, they go towards the same goal with the direction slightly changing as time advances.

\begin{figure}
    \centering
    \includegraphics[width=1\columnwidth]{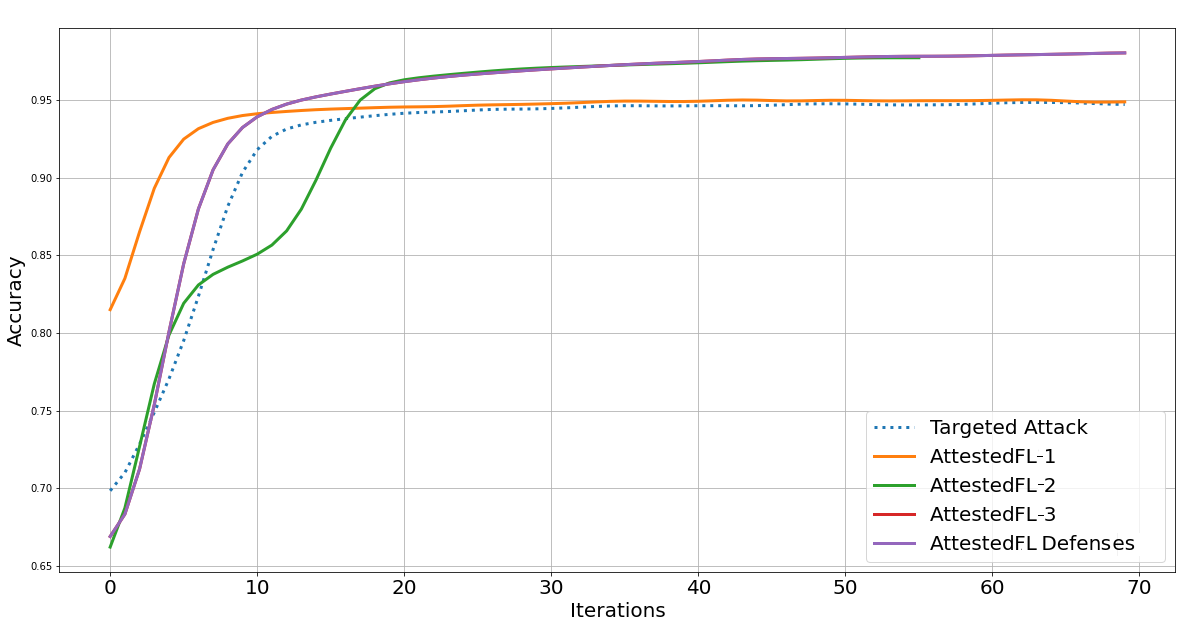}
    \caption{Accuracy of the global model at every training iteration of the different layers of attestedFL when defending against a targeted attack on the MNIST model.}
    \label{fig:targeted}
\end{figure}

Since untargeted attacks reduce the overall performance of the primary task, they are easier to detect. On the other hand, targeted attacks are harder to detect as the goal of the adversary is often unknown a priori. However, Figure~\ref{fig:targeted} shows the performance at every training iteration of the different layers of attestedFL when defending against a targeted attack on MNIST. 

The results show that when all defense layers are activated, attestedFL is able to protect against the targeted attack because we notice that the accuracy increased compared to the under attack scenario. Particularly, attestedFL-1 increased the accuracy of only 1\%. Attackers were able to evade detection by attestedFL-1 because in the targeted scenario, as time advances, attackers are training, meaning that they are advancing towards the same goal. The flags of attestedFL aim at detecting workers that are not training, thus the attackers were not flagged. On the other hand, attestedFL-2 performed better in this scenario. As training advanced, the attacker was sending local model updates that were less and less correlated between each other. Since the attacker was not able to maintain cosine similarity of successive local model updates throughout the training, the attacker was flagged by attestedFL-2. attestedFL-3 also enabled the detection of the attacker at every iteration because the performance on the quasi-validation set at the chief node decreased over time. However, an attacker that has access to the validation set can train on it locally and evade detection by attestedFL-3.

\subsubsection{Scalability and computation overhead}

\begin{figure}
    \centering
    \includegraphics[width=1\columnwidth]{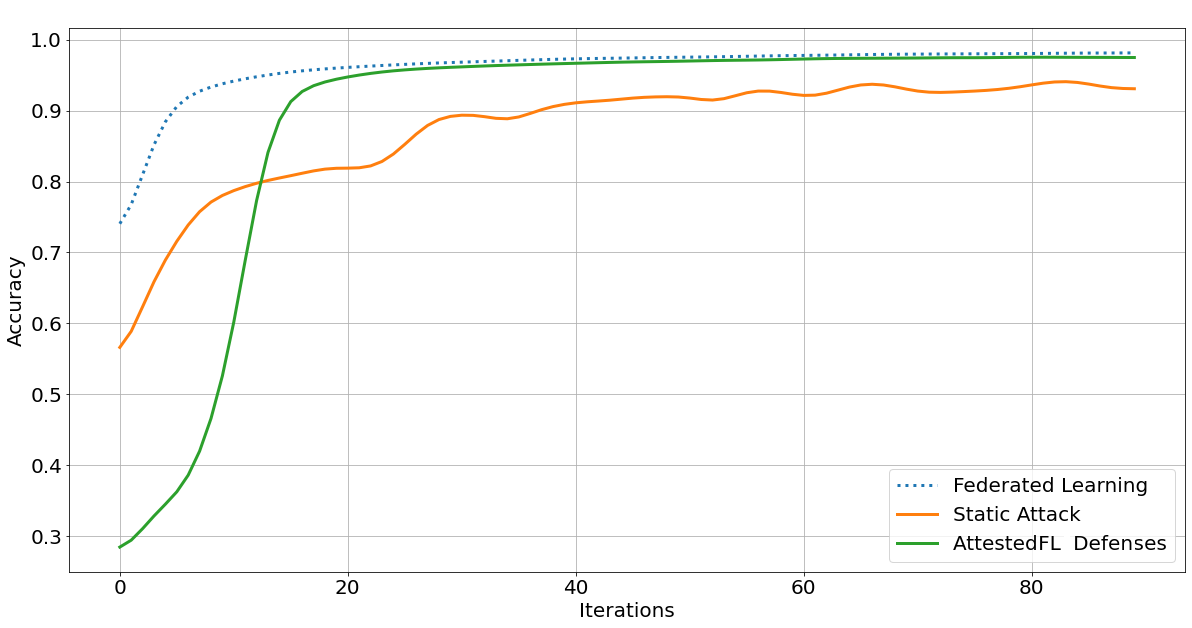}
    \caption{Effect of attestedFL against an attack on the MNIST model under different scales of federated learning settings.}
    \label{fig:MNIST_50nodes}
\end{figure}

In this section, we vary the number of \textit{workers} in the federated learning task and present the impact of the attack/defense in this scenario. Figure~\ref{fig:MNIST_50nodes} compares the accuracy of the global model at every iteration when attestedFL is protecting against an attack on MNIST when 50 \textit{workers} are involved in the task. attestedFL improved the accuracy of approximately 64.9\%. This demonstrates that attestedFL is scalable.

To evaluate the computation overhead incurred by our defense, we run the system with and without attestedFL with 10 – 50 \textit{workers} on a commodity CPU. We present the computation overhead incurred by training an MNIST and a CIFAR classifier. The attack in both settings is a static continuous untargeted poisoning on the FL process done by a fixed number of \textit{workers}.

\begin{figure}
    \centering
    \includegraphics[width=1\columnwidth]{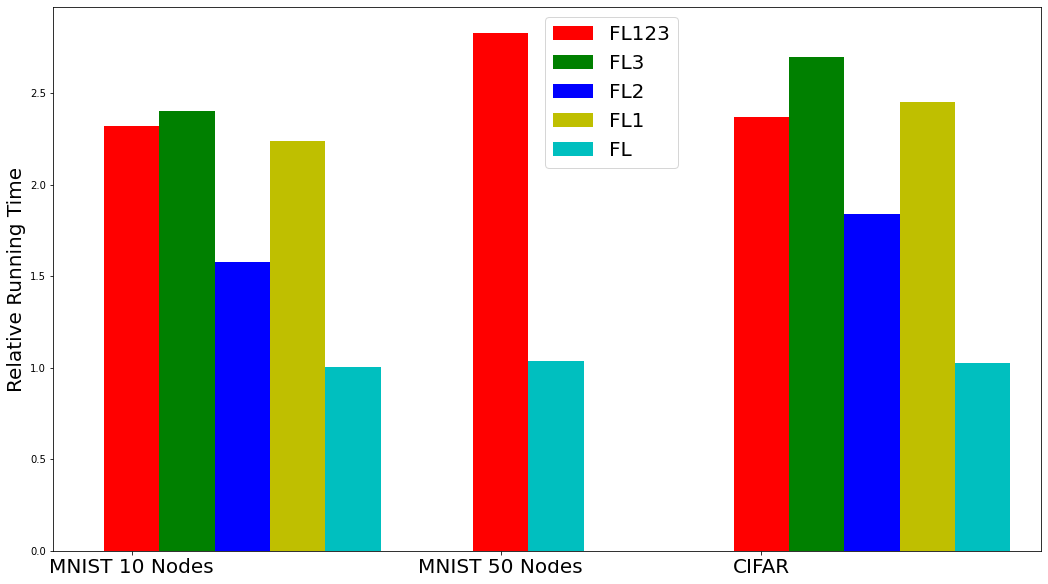}
    \caption{Running time overhead of attestedFL as compared to Federated Learning for MNIST and CIFAR with 10 – 50 \textit{workers} on a commodity CPU. The attack consists of a fixed number of compromised \textit{workers} that perform continuous untargeted poisoning during the entire learning process.}
    \label{fig:computation}
\end{figure}

Precisely, Figure~\ref{fig:computation} plots the relative slowdown added by attestedFL-1, attestedFL-2, attestedFL-3 and also when all the defenses are activated at the same time for the 10 and 50 \textit{workers} scenarios of the different training settings.

We notice that attestedFL-1 and attestedFL-3 are more expensive than attestedFL-2 in all scenarios and under different datasets. Moreover, when training a deep learning model, the cost of training is high enough that the relative slowdown from attestedFL when all the defences are activated at the same time is negligible than when attestedFL-1 or attestedFL-2 or attestedFL-3 are activated separately. Finally, the running time overhead added by the defense is more significant when we increased the number of \textit{workers} in the FL process. When 50 nodes are involved in the MNIST scenario, attestedFL took 2.8 more time to complete an iteration compared to the time it took to complete a round of FL training without the defense. As this was not the primary focus, our Python prototype is not optimized, but there are known optimizations to improve the speed of computing distances and cosine similarities at scale.

In our implementation, the \textit{chief} does not need to replicate all local model updates sent by each \textit{worker} in order to maintain their history. Only the previous one needs to be stored in order to compute the cosine similarity required in attestedFL-2, and on the next iteration. This model gets replaced for each \textit{worker}. In this way, history is maintained through a sliding window of a fixed-length for the previous distance values, cosine similarities and error metrics so that storage is optimised during training. We tested many implementations of our defense and some of them ran out of memory. The best optimisation was the use of this sliding window which represents a better assessment of the fact that the closer the values are to the current training step, the more the contribution is to the training. 

The machine learning community proposed the Multi-Krum algorithm aiming at preventing an attack by a single attacker \cite{blanchard2017machine} and FoolsGold prevents the targeted attack. Independently, these two defenses fail to defend both attacks concurrently, either by failing to detect the single attacker scenario or by allowing attackers to collude to overpower the system (against Multi-Krum). Our defense will not interfere with the above defenses as it adds a protection against the untargeted attack where the other two defenses fail. 

If the attacker can choose different sets of \textit{workers} to control at each epoch as in the randomized attack, which represents a more sophisticated attack, our defense forces it to conduct a bounded number of attacks per compromised \textit{worker} otherwise the attacker will be detected by attestedFL. The behavior of compromised nodes will be detected as malicious by our technique even with a defense-aware attacker. To evade detection by our technique, the attacker must try to exploit very large networks and launch attacks at large scale while trying to show a pattern in the training of the \textit{worker} and see if they can attain their goal of decreasing performance. 

\section{Conclusion}
\label{sec:conclussions}

We presented attestedFL, a defense against untargeted model poisoning attacks in federated learning that proved to reduce the attack effectiveness by monitoring through state persistence if the \textit{workers} in the model are training. While the attacks tried to decrease the accuracy, attestedFL increased it, showing the efficiency and security of our proposed defense. Our defense looks at a pattern in the training of the \textit{worker} in its subset of previously uploaded consecutive local model updates. attestedFL would force an attacker to choose from iteration to iteration weights inline with the training process expected at the local node, thus becoming less focused towards the poisoning objective. 


Our defense is mostly effective if staged by the \textit{chief} at every iteration. However, since the intuition is that for poisoning to take effect, recent local model updates of \textit{workers} have a higher weight on the global model, the assumption can be relaxed. In future work, we will explore such optimizations of our defense and evaluate it under different architectures and topologies (vertical FL, decentralized FL). We will also investigate subversion strategies. Given bounds on how much a deviation in a local model update can influence the output of the global model, an investigation is required into the relationships between the number of single shot attacks that must be done, the time to launch the attack, the size of the network and the impact required for a stealthy attack to succeed and circumvent attestedFL. 

The data used to support the findings of this study as well as the code are made publicly available by the authors on~\url{https://git.io/JtLzH}.
\bibliographystyle{IEEEtran}
\bibliography{ref_really}
\end{document}